\documentclass[aps,prd,preprint,groupedaddress,showpacs]{revtex4}

\usepackage{epsfig}
\usepackage{graphics}
\usepackage{slashed}
\usepackage{color}
\usepackage{multirow}
\begin{document}

\leftmargin -2cm
\def\choosen{\atopwithdelims..}

\boldmath
\title{High-Energy Factorization for Drell-Yan process in $pp$ and $p\bar{p}$ collisions
with new Unintegrated PDFs} \unboldmath

\author{\firstname{M.A.} \surname{Nefedov}} \email{nefedovma@gmail.com}
\affiliation{Samara National Research University, Moskovskoe Shosse,
34, 443086, Samara, Russia}

\author{\firstname{V.A.}\surname{Saleev}} \email{saleev@samsu.ru}

\affiliation{Samara National Research University, Moskovskoe Shosse,
34, 443086, Samara, Russia}

\affiliation{Joint Institute for Nuclear Research, Dubna, 141980
Russia}

\begin{abstract}
The formalism for uniform description of Drell-Yan transverse-momentum spectrum is presented in a framework of High-Energy Factorization, which smoothly interpolates between Collins-Soper-Sterman formalism at $|{\bf q}_T|\ll Q$ and usual Collinear Parton Model at $|{\bf q}_T|\sim Q\ll \sqrt{S}$. The new formula for deriving Unintegrated Parton Distribution Functions(UPDFs) from collinear ones is introduced, which leads to excellent description of the shape of $Z$-boson $|{\bf q}_T|$-spectrum at high energies up to $|{\bf q}_T|/\sqrt{S}\simeq 0.02$. Description of normalized $|{\bf q}_T|$-distributions at low energies is achieved via the fit of non-perturbative parameters of quark UPDFs. Reasonable description of angular distributions of leptons in the dilepton center-of-mass frame is also obtained with new UPDFs.
\end{abstract}



\maketitle

\section{Introduction}
\label{sec:Int} 
 
The transverse-momentum (${\bf q}_T$) distribution of Drell-Yan(DY) lepton pairs with large invariant mass $Q\gg\Lambda_{\rm QCD}$, produced in hadronic collisions, continues to attract a lot of attention from theorists and experimentalists alike. High-precision data on the $|{\bf q}_T|$-spectrum of lepton pairs with $Q$ close to the $Z$-boson mass had been obtained very recently by ATLAS Collaboration~\cite{ATLAS-2019} in $pp$-collisions with highest energy achieved so far, $\sqrt{S}=13$ TeV. Complimentary set of data on transverse-momentum distribution at lower values of $Q$ had been recently published by PHENIX Collaboration at RHIC Collider with $\sqrt{S}=200$ GeV~\cite{PHENIX}, which partially fills the gap between Drell-Yan data obtained in fixed-target experiments in 1980s and early 1990s~\cite{E-288,R-209,E-605} and data obtained at Tevatron~\cite{CDF-1999} and LHC energies. 

  From the theory side, the description of Drell-Yan $|{\bf q}_T|$-spectrum at $|{\bf q}_T|\ll Q$ have recently reached maturity, with the achievement~\cite{Scimemi:2019cmh,Bacchetta:2019sam} of Next$^{3}$-to-Leading Logarithmic (N$^3$LL) accuracy of the resummation of higher-order perturbative QCD corrections, enhanced by large $\ln Q^2/{\bf q}_T^2$, consistently interfaced with non-perturbative effects, important at $|{\bf q}_T|\sim\Lambda_{\rm QCD}$, in the context of the Transverse-Momentum Dependent(TMD)-factorization formalism~\cite{CollinsQCD}.  

  At the same time it has been observed~\cite{Bacchetta:2019tcu}, that Next-to-Leading Order calculation of the $|{\bf q}_T|$-spectrum in the {\it Collinear Parton Model(CPM)} of QCD can not describe normalization and shape of low-energy Drell-Yan data in the region $|{\bf q}_T|\gtrsim Q$, where $\ln Q^2/{\bf q}_T^2$-enhancement of higher-order corrections is not present, and fixed-order predictions should be applicable. The similar difficulty with the description of transverse-momentum spectrum of identified hadron in Semi-Inclusive Deep-Inelastic Scattering has been found in Ref.~\cite{Gonzalez-Hernandez:2018ipj}. The resummation of threshold effects up to Next-to-Leading Logarithmic Approximation improves the agreement with experimental data only marginally, as it has been shown in Ref.~\cite{Bacchetta:2019tcu}. In our opinion, this phenomenological puzzle is a manifestation of deeper theoretical issue with current formulation of TMD-factorization, which does not provide a unique prescription for the matching between TMD (the so-called $W$-term) and Collinear-factorization (the $Y$-term) parts of the calculation at $|{\bf q}_T|\simeq Q$ (see e.g. Ref.~\cite{Collins:2016hqq} for detailed discussion) and even lacks QED gauge-invariant definition for the $W$-term at $|{\bf q}_T|\sim Q$, see Refs.~\cite{Nefedov:2018vyt,Nefedov:2019hfn,Nefedov:2020bty}. 

  In the present paper, we approach the problem of uniform description of the $|{\bf q}_T|$-spectrum of Drell-Yan lepton pairs from a point of view of {\it High-Energy Factorization (HEF),} which initially has been introduced as a resummation tool for $\ln \hat{s}/(-\hat{t})$-enhanced corrections to the hard-scattering coefficients in Collinear Parton Model~\cite{Collins:1991ty,Catani:1994sq}, where invariants $\hat{s}$ and $\hat{t}$ refer to the partonic subprocess. Our {\it Parton Reggeization Approach (PRA)} is is a version of HEF, based on the {\it Modified Multi-Regge Kinematics (MMRK)} approximation for QCD scattering amplitudes. This approximation is accurate both in the Collinear limit, which drives the TMD-factorization and in the High-Energy (Multi-Regge) limit $\hat{s}\gg (-\hat{t})\sim {\bf q}_T^2\sim Q^2$ which is important for {\it Balitsky-Fadin-Kuraev-Lipatov(BFKL)}~\cite{BFKL1,BFKL2,BFKL3} resummation of $\ln \hat{s}/(-\hat{t})$-enhanced effects. This approximation allows us to derive the factorization formula for Drell-Yan cross-section, which is equivalent to the perturbative Collins-Soper-Sterman (CSS) formalism~\cite{Collins:1984kg} for $|{\bf q}_T|\ll Q$ and accuracy of which at $|{\bf q}_T|\sim Q$ is expected to increase power-like with decreasing values of $|{\bf q}_T|/\sqrt{S}$ and $Q/\sqrt{S}$. Thus, with increasing collision energy we should achieve a uniform description of $|{\bf q}_T|$-spectrum which does not require any dedicated matching procedure at $|{\bf q}_T|\simeq Q$.

  The present paper has the following structure. In the Sec.~\ref{sec:HEF} we introduce the MMRK approximation and derive factorization formula of PRA for the DY process; in the Sec.~\ref{sec:UPDF} we derive the Unintegrated Parton Distribution Function (UPDF) of PRA; in the Sec.~\ref{sec:PRA-CSS} we compare our cross-section formula at $|{\bf q}_T|\ll Q$ with the results of CSS formalism up to NLL; in the Sec.~\ref{sec:PRA-DY} we derive formulas for differential cross-section and squared LO PRA matrix element used in the numerical calculations; in the Sec.~\ref{sec:UPDF-fit} we compare our predictions with low-energy DY data and perform the fit of non-perturbative parameters of our UPDF; in the Sec.~\ref{sec:DY-Tev-LHC} we compare our predictions for normalized $|{\bf q}_T|$-spectra and coefficients parametrizing angular distributions of leptons in the center-of-mass frane of a lepton pair with High-Energy ATLAS~\cite{ATLAS-2019} and CDF data~\cite{CDF-1999} and in the Sec.~\ref{sec:concl} we summarize our conclusions.         

\section{PRA as High-Energy factorization at leading power}
\label{sec:HEF}
  The DY lepton-pair production at leading order in QED coupling constant $\alpha$ proceeds via the exchange of a virtual photon or $Z$-boson with four-momentum ($q$), thus the cross-section of this process, differential over invariant mass ($Q$, $q^2=Q^2$), rapidity ($y$) and transverse-momentum of the lepton pair, admits a well-known factorization into a convolution of leptonic ($L_{\mu\nu}$) and hadronic ($W_{\mu\nu}$) tensors. The latter can be written in the framework of CPM as follows:
\begin{equation}
W^{(a\bar{a})}_{\mu\nu} = \sum\limits_{i,j} \int\limits_{x_+}^1\frac{dz_+}{z_+} \tilde{f}_i\left( \frac{x_+}{z_+},\mu_F^2 \right) \int\limits_{x_-}^1\frac{dz_-}{z_-} \tilde{f}_j\left( \frac{x_-}{z_-},\mu_F^2 \right)\ w^{(ij,a\bar{a},{\rm CPM})}_{\mu\nu}\left(z_+,z_-\right),\label{eq:CPM-DY} 
\end{equation}
where $x_+=q_+/P_1^+=Q_Te^y/\sqrt{S}$, $x_-=q_-/P_2^-=Q_Te^{-y}/\sqrt{S}$, $Q_T^2=Q^2+{\bf q}_T^2$, $\tilde{f}_i(x,\mu^2)=xf_i(x,\mu^2)$ is the momentum-density PDF, indices $i,j=q,\bar{q},g$ run over parton species, $w_{\mu\nu}^{(ij,a\bar{a},{\rm CPM})}$ is the partonic tensor and index $a=V,A$($\bar{a}=V,A$) denote respectively the vector or axial-vector coupling of a vector boson to the quark line in the partonic amplitude (complex-conjugate amplitude). Here and below we use the Sudakov basis-vectors $n_-^\mu=2P_1^\mu/\sqrt{S}$ and $n_+^\mu=2P_2^\mu/\sqrt{S}$ where $S=2P_1P_2$ to define light-cone components of a four-momentum $k^\mu$ as $k_{\pm}=n_{\pm}k$.

 To isolate the $x_{\pm}$-dependence of the cross-section, one introduces the Mellin transform: 
\[
\tilde{f}_i(x,\mu^2)=\int\frac{dN}{2\pi i} x^N \tilde{f}^{(i)}_N(\mu^2),
\] 
and then the differential cross-section of a DY lepton pair production via virtual-photon exchange can be written as:
\begin{equation}
\frac{d\sigma}{dQ^2 d{\bf q}_T^2 dy} = \frac{\alpha}{3\pi Q^2 Q_T^4} \int\frac{dN_1 dN_2}{(2\pi i)^2}\ x_+^{N_1} x_-^{N_2} \times \tilde{f}^{(i)}_{N_1}(\mu_F^2) \tilde{f}^{(j)}_{N_2}(\mu_F^2) H^{(ij)}_{N_1,N_2}(p^2),\label{eq:CPM-DY-N}
\end{equation}
where $H^{(ij)}_{N_1,N_2}$ is a Mellin-transform of the dimensionless hard-scattering coefficient $H_{ij}(z_+,z_-,p^2)$:
\begin{equation}
H^{(ij)}_{N_1,N_2}(p^2)=\int\limits_0^1dz_+ dz_-\ z_+^{-N_1-1} z_-^{-N_2-1} H_{ij}(z_+,z_-,p^2), \label{eq:H-Mellin}
\end{equation}
 and we have introduced dimensionless parameter $p^2={\bf q}_T^2/Q_T^2$. Note that $0\leq p^2\leq 1$ and $p\to 0$ corresponds to the collinear regime with ${\bf q}_T^2\ll Q^2$, while $p=1$ corresponds to the production of the on-shell photon. 

  The analytic structure of integrand in Eq.(\ref{eq:CPM-DY-N}) at fixed order in $\alpha_s$ is well-known, see e.g.~\cite{IKLQCD}, sec. 2.8. The left-most poles in $N_{1,2}$ correspond to small-$x$ behaviour of PDFs and for most of existing PDF fits this poles have ${\rm Re}\ N_{1,2}\simeq -1/2$ for $\mu_F\gg 1$ GeV both for quark and gluon PDFs due to doubly-logarithmic asymptotics of  Dokshitzer-Gribov-Lipatov-Altarelli-Parisi (DGLAP) evolution~\cite{DGLAP1,DGLAP2,DGLAP3} of PDFs. Singularities of $H^{(ij)}_{N_1,N_2}$ have ${\rm Re}\ N_{1,2}\geq 0$. Therefore, to capture the leading-power $x_{\pm}$-dependence of the cross-section, one have to come-up with an accurate approximation for $H^{(ij)}_{N_1=-1/2,N_2=-1/2}(p)$.

  As a simple test-case let's consider the LO CPM coefficient function:
\begin{eqnarray}
H_{ij}^{\rm (LO)}(z_+,z_-,p^2)&=&z_+z_-\overline{|{\cal A}_{ij}|^2}\left( \frac{\hat{t}}{\hat{s}}=1-z_+,\frac{Q^2}{\hat{s}}=(1-p^2)z_+z_- \right) \nonumber \\
&\times& \delta\left( \frac{1-z_+}{z_+z_-} - \frac{p^2}{1-z_-} \right), \label{eq:H-ex}
\end{eqnarray}
where $\overline{|{\cal A}_{ij}|^2}$ are the well-known squared matrix elements of $2\to 2$ partonic subprocesses $q(k_1)+\bar{q}(k_2)\to \gamma^*(q) + g(k_3)$ and $q(k_1)+g(k_2)\to \gamma^*(q) + q(k_3)$ respectively, averaged over color and spin quantum numbers of initial-state partons:
\begin{eqnarray*}
\overline{|{\cal A}_{q\bar{q}}|^2}&=&(4\pi)^2\alpha\alpha_se_q^2\frac{C_F}{N_c}\frac{2Q^2\hat{s}+\hat{t}^2+\hat{u}^2}{\hat{t}\hat{u}}, \\
\overline{|{\cal A}_{qg}|^2}&=&(4\pi)^2\alpha\alpha_se_q^2\frac{T_R}{N_c}\frac{(Q^2+\hat{t})^2+(Q^2-\hat{s})^2}{-\hat{t}\hat{s}},
\end{eqnarray*} 
where $C_F=(N_c^2-1)/2N_c$, $T_R=1/2$, $\hat{s}=(k_1+k_2)^2$, $\hat{t}=(k_1-q)^2$ and $\hat{u}=(k_2-q)^2$.

  Many of the standard ``DGLAP'' Parton-Showers Monte-Carlo event generators, such as {\it e.g.} \texttt{PYTHIA}~\cite{Sjostrand:2006za}, are based on the collinear approximation for matrix elements with additional emissions, which is accurate in the limit ${\bf q}_T^2\sim(-\hat{t})\ll Q^2$, see e.g. Eq.~(4.9) of Ref~\cite{Catani:1996vz}. In this limit, the amplitude of the $2\to 2$ process factorizes into a product of $q(k_1)+\bar{q}(\tilde{q}_2)\to\gamma^*(q)$ amplitude with on-shell kinematics: $\tilde{q}^2_2=0$, which requires $\tilde{q}_2^+\tilde{q}_2^--{\bf q}_{T2}^2=0$, and the factor describing the ``splitting'' of a parton $i(k_2)\to j(k_3)+\bar{q}(q_2)$:
\begin{equation}
\overline{|{\cal A}_{qi}|^2}=\overline{|{\cal A}(q(k_1)+\bar{q}(\tilde{q}_2)\to \gamma^*(q))|^2}\times (8\pi\alpha_s) \frac{P_{\bar{q}i}(z_-)}{(-\hat{t})z_-},\label{eq:coll-approx}
\end{equation}
where $i=g,\bar{q}$, $P_{\bar{q}\bar{q}}(z)=C_F(1+z^2)/(1-z)$, $P_{\bar{q}g}(z)=T_R(z^2+(1-z)^2)$ are non-regularized DGLAP splitting functions and
$\overline{|{\cal A}(q(k_1)+\bar{q}(\tilde{q}_2)\to \gamma^*(q))|^2}= 4\pi \alpha Q^2/N_c$.

\begin{figure}
\begin{center}
\includegraphics[width=0.2\textwidth]{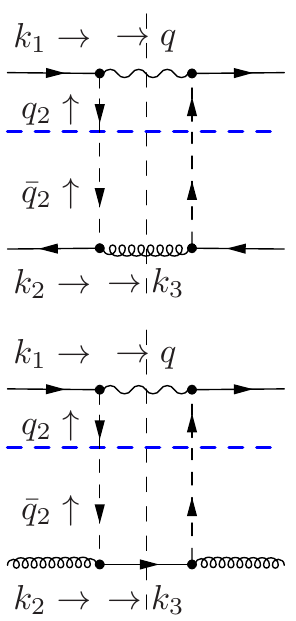}\hspace{1cm}
\includegraphics[width=0.2\textwidth]{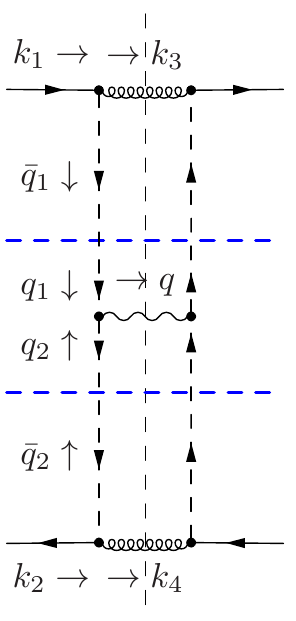}
\end{center}
\caption{Diagrammatic representations of squared (M)MRK amplitudes for $q+\bar{q}\to \gamma^*+g$ (upper diagram on the left panel), $q+g\to \gamma^*+q$ (lower diagram on the left panel) and $q+\bar{q}\to \gamma^* + 2g$ (right panel) processes. Dashed lines denote Reggeized quarks, solid dots denote Fadin-Sherman vertices. The ``small'' light-cone momentum components $\bar{q}_1^-$ and $\bar{q}_2^+$ are neglected beyond thick dashed lines.\label{fig:MRK_2-2}}
\end{figure}

 Another kinematic limit, in which QCD amplitudes admit simple factorization, is the Regge limit $z_-\to 0$, while relation between $Q^2$ and ${\bf q}_T^2$ can be arbitrary. In this {\it Multi-Regge Kinematics (MRK)}, final-state particles are highly-separated in rapidity. Asymptotic expression for QCD amplitudes with quark-exchange in $\hat{t}$-channel in this limit can be obtained using the formalism of Gauge-Invariant EFT for Multi-Regge processes in QCD~\cite{Lipatov95,LV}. For both considered squared amplitudes, this asymptotics depicted diagrammatically in the left panel of the Fig.~\ref{fig:MRK_2-2}, can be written as:
\begin{equation}
\overline{|{\cal A}_{qi}^{(z_-\ll 1)}|^2}= \frac{(4\pi\alpha)}{2N_c}{\cal P}^{\mu\nu}(q)\times {\rm tr}\left[ \Gamma^{(+)}_\mu(-q,q_2) \hat{k}_1 \Gamma^{(+)}_\nu(-q,q_2) \hat{S}^{(-)}_{\bar{q}i}(k_2,\bar{q}_2) \right], \label{eq:M_qi_MMRK-0}
\end{equation}  
where $\hat{k}=k_\mu\gamma^\mu$, $q_2=q-k_1$, $\bar{q}_2=k_2-k_3$, ${\cal P}_{\mu\nu}(q)=-g_{\mu\nu}+q_\mu q_\nu/Q^2$ is the polarization sum for off-shell photon, factor $1/2$ corresponds to the averaging over helicities of initial-state quark, Fadin-Sherman scattering vertices~\cite{FadinSherman76, FadinSherman77} are:
\begin{equation}
\Gamma^{(\pm)}_\mu(k,p)=\gamma_\mu+\hat{p}\frac{n^\pm_\mu}{k^\pm}, \label{eq:F-S_scatt}
\end{equation}
and factors $\hat{S}^{(-)}_{\bar{q}i}(k_2,\bar{q}_2)=\hat{S}^{(-)}_{qi}(k_2,\bar{q}_2)$ correspond to the lower part of diagrams in the Fig.~\ref{fig:MRK_2-2}, with the following general expressions for $\hat{S}^{(\pm)}_{qi}$:
\begin{eqnarray}
\hat{S}^{(\pm)}_{qq}(k,p)&=& (4\pi \alpha_s) C_F  \frac{1}{2} \hat{P}_{\pm}\frac{i\hat{p}}{p^2}\Gamma^{(\pm)}_\rho(p-k,-p)\hat{k}\Gamma^{(\pm)}(p-k,-p)\frac{(-i)\hat{p}}{p^2}\hat{P}_{\mp}\times P^{\rho\sigma}(p-k), \label{eq:S_qq-0} \\
\hat{S}^{(\pm)}_{qg}(k,p)&=& (4\pi \alpha_s) T_R \hat{P}_{\pm}\frac{i\hat{p}}{p^2}\Gamma^{(\pm)}_\rho(k,-p)(\hat{p}-\hat{k})\Gamma^{(\pm)}(k,-p)\frac{(-i)\hat{p}}{p^2}\hat{P}_{\mp}\times \frac{1}{2} P^{\rho\sigma}(k), \label{eq:S_qg-0}
\end{eqnarray}
where factors $1/2$ correspond to the averaging over helicities of initial-state quark or gluon, Dirac projectors $\hat{P}_{\pm}=\hat{n}_{\mp}\hat{n}_{\pm}/4$ are required by EFT Feynman rules~\cite{LV} and $P_{\mu\nu}(k)=-g_{\mu\nu}+(k_\mu n_\nu + k_\nu n_\mu)/(kn)-k_\mu k_\nu n^2/(kn)^2$ is the gluon polarization sum in general axial gauge. Note, that due to the structure of vertices (\ref{eq:F-S_scatt}) and conditions $k_2^2=0$ or $(q_2-k_2)^2=0$, the splitting-factors $\hat{S}^{(\pm)}_{qi}$ are invariant w.r.t. the choice of gauge-vector $n_\mu$. 

  In the Regge limit $z_-\ll 1$, light-cone components of $\bar{q}_2$ obey the hierarchy: $\bar{q}_2^+\ll 
\bar{q}_2^-=z_-k_2^-$ and ``small'' $\bar{q}_2^+$-component is usually neglected in the simplification of $\hat{S}^{(+)}_{\bar{q}i}(k_2,\bar{q}_2)$. However, this kinematic approximation is not necessary in the case of amplitudes with quark exchange in $\hat{t}$-channel, because relaxing it does not violate gauge-invariance of the splitting-factors. One can recover the $\bar{q}_2^+$ momentum component form the on-shell condition $(k_2-\bar{q}_2)^2=0$:
\[
\bar{q}_2^\mu= \frac{1}{2}\left( k_2^- z_- n_+^\mu - \frac{{\bf q}_{T2}^2n_-^\mu}{k_2^-(1-z_-)} \right) + q_{T2}^\mu,
\]
where we take into account that ${\bf q}_{T2}^2={\bf k}^2_{T3}={\bf q}^2_T$ and one finds that $\bar{q}_2^2=-{\bf q}_{T2}^2/(1-z_-)$. Substituting the latter approximation for $q_2$ into Eqns. (\ref{eq:S_qq-0}) and (\ref{eq:S_qg-0}) one obtains:
\begin{equation}
 \hat{S}^{(-)}_{qi}= 8\pi\alpha_s\frac{P_{\bar{q}i}(z_-)}{(-z_-\bar{q}_2^2)} \times \frac{1}{2} \left(\frac{\hat{n}_+k_2^- z_-}{2} \right),
\end{equation}
for both cases $\hat{S}^{(-)}_{qq}$ and $\hat{S}^{(-)}_{qg}$.  Substituting this result into Eq. (\ref{eq:M_qi_MMRK-0}) and calculating the trace:
\[
{\cal P}^{\mu\nu}(q)\ {\rm tr}\left[ \Gamma^{(-)}_\mu(q-k_1,-q) \hat{k}_1 \Gamma^{(-)}_\nu(q-k_1,-q)  \left(\frac{\hat{n}_+k_2^- z_-}{2} \right) \right] = 4Q_T^2,
\]
one obtains the {\it Modified MRK Approximation (MMRK)} for the considered squared amplitudes:
\begin{equation}
\overline{|{\cal A}_{qi}^{\rm (MMRK)}|^2}= \frac{4\pi\alpha}{N_c}Q_T^2 \times (8\pi\alpha_s) \frac{P_{\bar{q}i}(z_-)}{(-z_-\bar{q}_2^2)}.\label{eq:MMRK-approx}
\end{equation}
  Note that in the MMRK approximation, we have neglected the $\bar{q}_2^+$ light-cone component in the ``hard process'' (virtual photon production vertex in the left panel of Fig.~\ref{fig:MRK_2-2}), while keeping it in the calculation of the $\hat{S}^{(-)}_{qi}$. This approximation is more general than Eq.~(\ref{eq:coll-approx}), because it is accurate in two limits: ${\bf q}_T^2\ll Q^2$ for any $z_-$, and $z_-\ll 1$ for any hierarchy between $Q^2$ and ${\bf q}_{T}^2$. The MMRK analog of Eq.~(\ref{eq:H-ex}) reads:
\[
H_{qi}^{\rm (LO, MMRK)}(z_+,z_-,p^2)=z_-^2 \overline{|{\cal A}_{qi}^{\rm (MMRK)}|^2} \delta(z_+-1) \theta\left( \Delta({\bf q}_T^2,Q_T^2)-z_- \right),
\]  
where $\Delta(t,\mu)=\sqrt{\mu}/(\sqrt{t}+\sqrt{\mu})$ -- the Kimber-Martin-Ryskin-Watt cutoff function~\cite{Kimber:2001sc,Watt:2003mx,Watt:2003vf,Martin:2009ii}. The $\theta$-function in the last equation defines the region of applicability of MMRK-approximation to be only the case when rapidity of a virtual photon is larger than the rapidity of a quark or gluon. Indeed, the rapidity of a photon is $y_{\gamma^*}=\ln(q^+/q^-)/2=\ln(Q_T/(k_2^- z_-))$, while the rapidity of a final-state parton is $y_3=\ln(k_3^+/k_3^-)/2=\ln(|{\bf q}_T|/(k_2^-(1-z_-)))$, hence the condition $z_-<\Delta({\bf q}_T^2,Q_T^2)$ is equivalent to $y_{\gamma^*}>y_3$. In the opposite case, the ``$\hat{u}$-channel'' MMRK approximation should be used, which is obtained from approximation above by the replacement $z_+\leftrightarrow z_-$.

  The idea behind MMRK-approximation is not new. It was first successfully applied in the High-Energy Jets approach~\cite{HEJ1,HEJ2} where a good $\hat{t}$-channel-factorized approximation for QCD amplitudes with emissions of multiple additional partons has been constructed via relaxing of some kinematic constraints in corresponding MRK-asymptotic amplitudes, while preserving their QCD gauge-invariance. Later, the TMD-generalizations of usual DGLAP splitting functions describing the splitting of off-shell $\hat{t}$-channel partons have been constructed in Refs.~\cite{Hautmann:2012sh,Gituliar:2015agu,Hentschinski:2017ayz} using the same guiding principles. And recently it has been shown in Ref.~\cite{Nefedov:2020ecb}, that the problem of large NLO corrections for gluon-induced observables in HEF can be solved, if the MMRK approximation for QCD amplitudes is used to construct the UPDF evolution equation and corresponding double-counting subtraction terms at NLO.

\begin{figure}
\begin{center}
\includegraphics[width=0.45\textwidth]{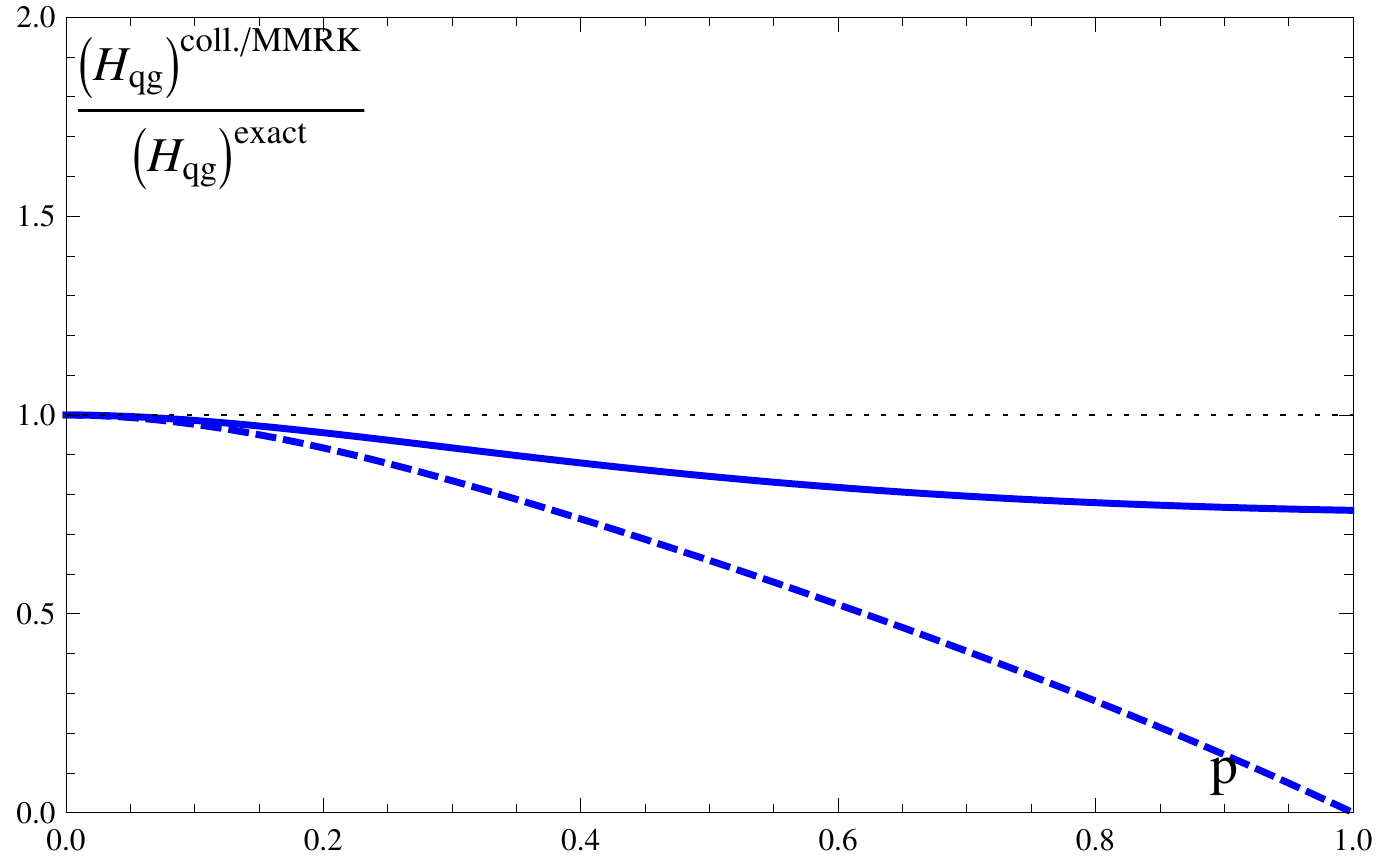} \includegraphics[width=0.45\textwidth]{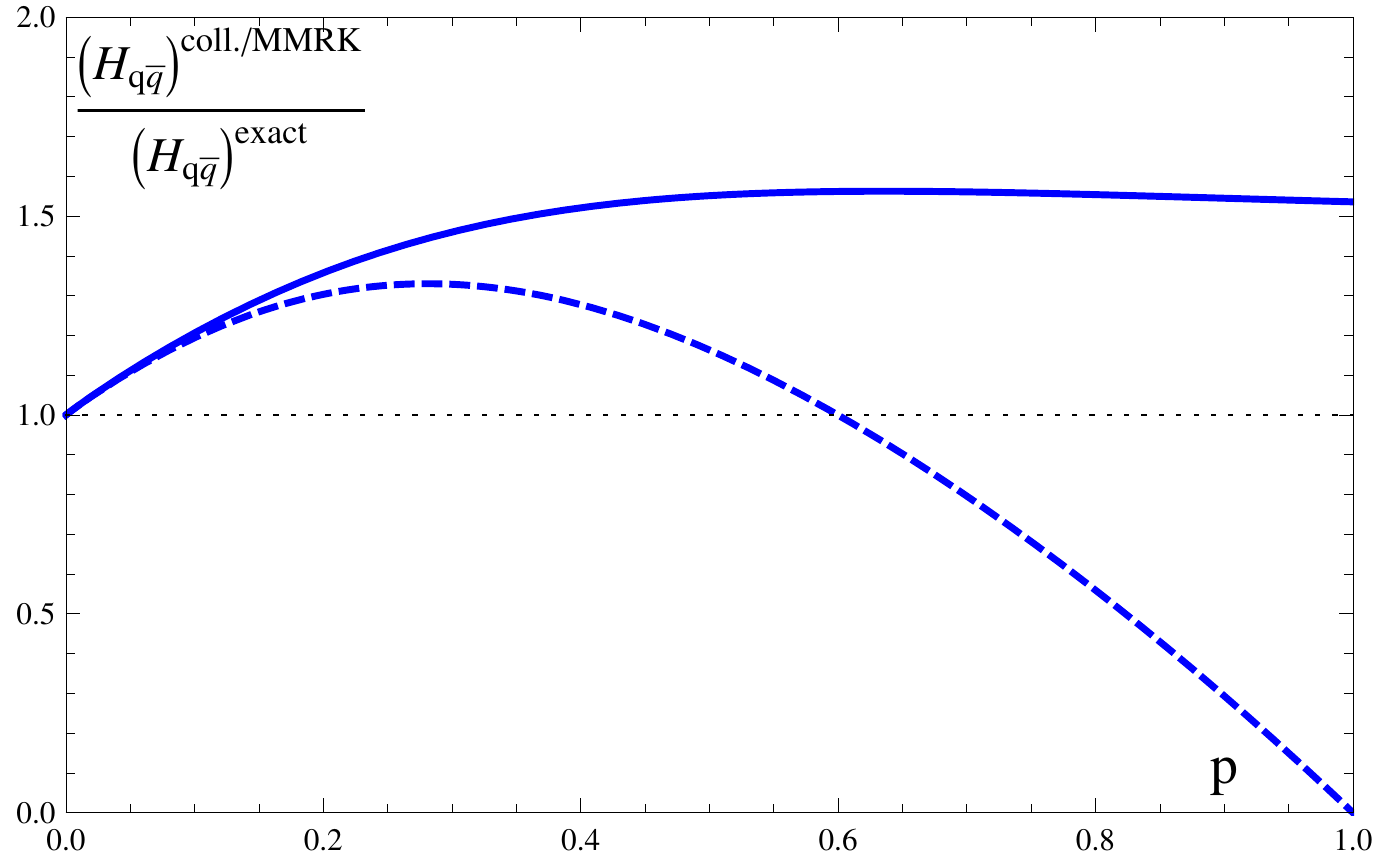}
\end{center}
\caption{ Ratio plots of the quantities $H^{(qg)}_{N_1=-1/2,N_2=-1/2}(p^2)$ (left panel) and $H^{(q\bar{q})}_{N_1=-1/2,N_2=-1/2}(p^2)$ (right panel) in the collinear (dashed curves) and MMRK (solid curves) approximations to the corresponding exact results, obtained with the use of Eqns.~(\ref{eq:H-Mellin}) and (\ref{eq:H-ex}), as functions of $p=|{\bf q}_T|/\sqrt{Q^2+{\bf q}_T^2}$.  \label{fig:H_qq_qg}}
\end{figure}

  In the Fig.~\ref{fig:H_qq_qg} we compare the functions $H^{(q\bar{q})}_{N_1=-1/2,N_2=-1/2}(p^2)$ and $H^{(qg)}_{N_1=-1/2,N_2=-1/2}(p^2)$ for collinear (\ref{eq:coll-approx}) and MMRK (\ref{eq:MMRK-approx}) approximations with corresponding exact result obtained by substitution of Eq.~(\ref{eq:H-ex}) into the Eq.~(\ref{eq:H-Mellin}). One can see, that MMRK approximation provides a reasonable estimate for $H^{(ij)}_{N_1=-1/2,N_2=-1/2}(p^2)$ up to $p\simeq 1/2$, i.e. for $|{\bf q}_T|<Q/\sqrt{3}$, while for larger values of $|{\bf q}_T|$ the error of MMRK-approximation reaches several tens of percent while staying flat all the way up to $p=1$. In contrast to this, the error of collinear approximation rapidly increases when $p\to 1$. Thus, using MMRK-approximation, one can construct the expression for the Drell-Yan $|{\bf q}_T|$-spectrum with effects of initial-state radiation factorized, which will capture the leading-power in $x_{\pm}$-dependence of the cross-section at least up to $|{\bf q}_T|\lesssim 0.6Q$.  

  Thanks to $\hat{t}$-channel-factorized nature of MMRK-approximation and process-independence of splitting-factors (\ref{eq:S_qq-0}) and (\ref{eq:S_qg-0}) one can derive the factorizaiton formula for the contribution of $2\to 3$ partonic process 
\begin{equation}
i(k_1)+j(k_2)\to \gamma^*(q) + i'(k_3) + j'(k_4),\label{eq:aux_2-3}
\end{equation}
with $i,j,i',j'=q,\bar{q},g$, to the hadronic tensor in CPM (\ref{eq:CPM-DY}). The MMRK approximation for one of such contributions is depicted diagrammatically on the right panel of the Fig.~\ref{fig:H_qq_qg} and for general subprocess of the type (\ref{eq:aux_2-3}) the MMRK partonic tensor in Eq.~(\ref{eq:CPM-DY}), integrated over phase-space of momenta $k_3$ and $k_4$ with $k_{3,4}^2=k_{3,4}^+k_{3,4}^--{\bf k}_{T3,4}^2=0$ can be written as:
\begin{eqnarray}
w_{\mu\nu}^{(ij,a\bar{a},{\rm CPM})}&=&\int\limits_0^{+\infty} \frac{dk_3^+ dk_4^-}{4k_3^+ k_4^-} \int \frac{d^2{\bf k}_{T3} d^2{\bf k}_{T4}}{(2\pi)^6} \int\limits_{-\infty}^{+\infty} d\bar{q}_1^- d\bar{q}_2^+ \int dq_1^+ d^2{\bf q}_{T1} \int dq_2^- d^2{\bf q}_{T2} \nonumber \\
&\times& \delta(k_1^+-k_3^+-q_1^+)\delta\left(\bar{q}_1^- + \frac{{\bf k}_{T3}^2}{k_3^+}\right) \delta(k_2^--k_4^--q_2^-)\delta\left( \bar{q}_2^+ + \frac{{\bf k}_{T4}^2}{k_4^-} \right) \nonumber \\
&\times& \delta^{(2)}({\bf k}_{T3}+{\bf q}_{T1}) \delta^{(2)}({\bf k}_{T4}+{\bf q}_{T2}) \times (2\pi)^4 \delta^{(4)}(q_1+q_2-q) \frac{{\cal A}_{\mu\nu}^{(ij,a\bar{a})}}{I(z_+,z_-)}, \label{eq:w-part-CPM}
\end{eqnarray}
where we have introduced integrations over light-cone components of $\hat{t}$-channel momenta $q_1$, $q_2$, as well as over $\bar{q}_1^-$, $\bar{q}_2^+$ and $I(z_+,z_-)=2Sx_+x_-/(z_+z_-)$ is the usual flux-factor of initial-state partons in CPM factorization formula (\ref{eq:CPM-DY}). The MMRK approximation for squared amplitude in Eq.~(\ref{eq:w-part-CPM}) reads:
\begin{eqnarray}
{\cal A}_{\mu\nu}^{(ij,a\bar{a})} &=& \sum\limits_{k,l} \frac{4\pi\alpha}{4N_c}C_{k}^{(a)}C_{\bar{l}}^{(\bar{a})} {\rm tr}\left[\hat{S}_{lj}^{(-)}(k_2,\bar{q}_2)  \Gamma_\mu^{(+-)}(q_1,q_2)(\delta_{aV}+\delta_{aA}\gamma_5) \right. \nonumber \\ 
&\times&\left. \hat{S}_{ki}^{(+)}(k_1,\bar{q}_1) (\delta_{\bar{a}V}-\delta_{\bar{a}A}\gamma_5)\Gamma_\nu^{(+-)}(q_1,q_2) \right]\delta_{k\bar{l}} \nonumber \\
&=& \sum\limits_{k,l} (8\pi\alpha_s)^2 \frac{P_{ki}(z_+) P_{lj}(z_-)}{z_+z_-\bar{q}_1^2 \bar{q}_2^2} \times w_{\mu\nu}^{(kl,a\bar{a})}, \label{eq:A_mn}
\end{eqnarray}   
where $\bar{q}_1^2=-{\bf q}_{T1}^2/(1-z_+)$, $\bar{q}_2^2=-{\bf q}_{T2}^2/(1-z_-)$ and HEF partonic tensors $w_{\mu\nu}^{(kl,a\bar{a})}$ are expressed in terms of $Q_+(q_1)+\bar{Q}_-(q_2)\to \gamma^*(q)$ Fadin-Sherman vertices~\cite{FadinSherman76, FadinSherman77} which we have applied to the case of Drell-Yan process for the first time in Ref.~\cite{Nefedov:2012cq}:
\begin{equation}
\Gamma^{(+-)}_\mu(q_1,q_2)=\gamma_\mu-\hat{q}_1\frac{n_\mu^-}{q_2^-} - \hat{q}_2\frac{n_\mu^+}{q_1^+}, \label{eq:F-S_QQ-gamma}
\end{equation}
as follows:
\begin{eqnarray}
w_{\mu\nu}^{(kl,a\bar{a})}&=&\delta_{k\bar{l}} \frac{(4\pi\alpha)}{4N_c} C_{k}^{(a)}C_{\bar{l}}^{(\bar{a})}{\rm tr}\left[ \left( \frac{q_1^+}{2}\hat{n}_- \right)\Gamma_\mu^{(+-)}(q_1,q_2) (\delta_{aV}+\delta_{aA}\gamma_5) \right. \nonumber \\
&\times&\left.  \left( \frac{q_2^-}{2} \hat{n}_+ \right) (\delta_{\bar{a}V}-\delta_{\bar{a}A}\gamma_5) \Gamma_\nu^{(+-)}(q_1,q_2)    \right], \label{eq:w-PRA}
\end{eqnarray}
with $C_q^{(a)}=\delta_{aV}e_q$ for the photon-quark coupling and 
\begin{equation}
C_q^{(a)}=\frac{1}{\sin (2\theta_W)} \left( (\delta_{aV}-\delta_{aA}) T_z^{(q)} -2 \delta_{aV} e_{q}\sin^2\theta_W  \right),\label{eq:C-fact}
\end{equation}
for the $q\bar{q}Z$ or $l^+l^-Z$-coupling, where isospin projection $T_z^{(q)}=+1/2$($-1/2$) for up-type(down-type) quarks, $T_z^{(l)}=-1/2$ for charged leptons and lepton/quark charges $e_{q/l}$ are taken in units of positron charge.  In our numerical calculations we adopt the following numerical values for $M_Z=91.1876$ GeV, $\Gamma_Z=2.4952$ GeV and $\sin^2\theta_W=0.2314$~\cite{10.1093/ptep/ptaa104}.

  Expression (\ref{eq:w-PRA}) for partonic tensor in HEF is free from any gauge ambiguities at ${\bf q}_T^2\sim Q^2$, since it exactly satisfies the Ward identity $q^\mu w_{\mu\nu}^{(kl,a\bar{a})} = q^\nu w_{\mu\nu}^{(kl,a\bar{a})} =0$, see also~\cite{Nefedov:2018vyt, Nefedov:2019hfn}.

  To complete the derivation of the HEF factorization formula one substitutes Eq.~(\ref{eq:A_mn}) to Eq.~(\ref{eq:w-part-CPM}), integrates-out $k_3^+$, $k_4^-$, $\bar{q}_1^-$, $\bar{q}_2^+$ and ${\bf k}_{T3,4}$ using corresponding delta-functions, then introduces momentum-fractions $x_1=q_1^+/P_1^+$ and $x_2=q_2^-/P_2^-$ instead of $q_1^+$ and $q_2^-$ and finally substitutes the result for $w_{\mu\nu}^{(ij,a\bar{a})}$ into Eq.~(\ref{eq:CPM-DY}) to obtain:
\begin{eqnarray}
W_{\mu\nu}&=&\sum\limits_{k,l}\int\limits_{0}^1\frac{dx_1}{x_1} \int\frac{d^2{\bf q}_{T1}}{\pi} \Phi^{\rm (tree-level)}_{k}(x_1,{\bf q}_{T1},\mu_Y) \int\limits_{0}^1\frac{dx_2}{x_2} \int\frac{d^2{\bf q}_{T2}}{\pi} \Phi^{\rm (tree-level)}_{l}(x_2,{\bf q}_{T2},\mu_Y) \nonumber \\
&\times & (2\pi)^4 \delta(q_1+q_2-q) \frac{w_{\mu\nu}^{(kl,a\bar{a})}}{2Sx_1x_2}, \label{eq:Wmn-PRA}
\end{eqnarray} 
  where the tree-level {\it unintegrated PDFs (UPDFs)} are :
\begin{equation}
  \Phi^{\rm (tree-level)}_i(x,t,\mu_Y^2)= \frac{\alpha_s(\mu_R)}{2\pi} \frac{1}{t} \sum\limits_{j=q,\bar{q},g}\int\limits_x^1 dz\ P_{ij}(z) \tilde{f}_j\left( \frac{x}{z}, \mu_F^2 \right) \theta\left( \Delta(t,\mu_Y^2)-z \right). \label{eq:UPDF-tree}
\end{equation}

 The $\theta$-functions in Eq.(\ref{eq:UPDF-tree}) enforce the rapidity-ordering between particles in the final-state $y_3>y_{\gamma^*}>y_4$, for our MMRK approximation for the squared amplitude and kinematics to be applicable.  The natural choice of {\it rapidity scale} $\mu_Y$ for the case of Drell-Yan process is $\mu_Y\sim Q_T$.  As it follows from the discussion above, Eq.~(\ref{eq:UPDF-tree}) is accurate in the region $\mu_Y\sim\mu_F\sim\mu_R\sim t$ with $x_{\pm}\ll z_{\pm}\ll 1$. For $t\ll \mu_Y$ the tree-level UPDF contains a collinear divergence $\sim 1/t$ signaling the break-down of fixed-order perturbation theory for this object.

An important feature of Eq.~(\ref{eq:Wmn-PRA}), which is critical in the region ${\bf q}_T^2\sim Q^2\ll S$ is, that flux-factor $2Sx_1x_2$ is used for off-shell initial-state partons with $q_1^2=-{\bf q}_{T1}^2<0$ and $q_2^2=-{\bf q}_{T2}^2<0$. This prescription follows from the derivation of Eqns.~(\ref{eq:Wmn-PRA}) and (\ref{eq:UPDF-tree}), presented above. The similar derivation for gluon-induced processes has been given in Sec. II of our Ref.~\cite{NS_PRA}. We stress again, that this prescription is necessary for consistency of the cross-section formula of High-Energy Factorization with exact QCD results in Regge limits with $x_+\ll z_+\ll 1$ and/or $x_-\ll z_-\ll 1$, which give a major contribution to the cross-section in the regime ${\bf q}_T^2\sim Q^2\ll S$. Thus the prescription of Eq.~(\ref{eq:Wmn-PRA}) for the flux-factor should be used consistently with the gauge-invariant amplitudes based on the vertex (\ref{eq:F-S_QQ-gamma}) and both of this factors are important for the $|{\bf q}_T|$-distribution at $|{\bf q}_T|\sim Q$. In connection with this we would like to emphasize that only the ``off-shell cross-section'' formula (56) in recent Ref.~\cite{Golec-Biernat:2019scr}, is self-consistent at $|{\bf q}_T|\sim Q$, while the ``on-shell'' cross-section formula (47) is applicable only for $|{\bf q}_T|\ll Q$. 

\section{Unintegrated PDF with exact normalization}
\label{sec:UPDF}
 To resolve a divergence problem of Eq.~(\ref{eq:UPDF-tree}) we follow the standard definition of the UPDF in BFKL formalism (see e.g. Eq. (2.4) in the Ref.~\cite{Collins:1991ty} or Sec. 1 in~\cite{Kotikov:2002ab}) and require that:
\begin{equation}
\int\limits_0^{\mu^2} dt\ \Phi_i(x,t,\mu^2) = \tilde{f}_i(x,\mu^2), \label{eq:norm-cond}
\end{equation}
which is equivelent to:
\begin{equation}
\Phi_i(x,t,\mu^2)=\frac{d}{dt}\left[ T_i(t,\mu^2,x) \tilde{f}_i(x,t) \right], \label{eq:UPDF-deriv-form}
\end{equation}
with some function $T_i(t,\mu^2,x)$ which is usually referred to as {\it Sudakov formfactor}, satisfying the boundary conditions $T_i(t=0,\mu^2,x)=0$ and $T_i(t=\mu^2,\mu^2,x)=1$. We will obtain the latter by multiplying Eq.~(\ref{eq:UPDF-tree}) on the formfactor:
\begin{equation}
  \Phi_i(x,t,\mu_Y^2)= \frac{\alpha_s(t)}{2\pi} \frac{T_i(t,\mu^2,x)}{t} \sum\limits_{j=q,\bar{q},g}\int\limits_x^1 dz\ P_{ij}(z) \tilde{f}_j\left( \frac{x}{z}, t \right) \theta\left( \Delta(t,\mu_Y^2)-z \right), \label{eq:UPDF}
\end{equation}
and asking for {\it exact equivalence} of two definitions (\ref{eq:UPDF-deriv-form}) and (\ref{eq:UPDF}). Note, that Eq. (\ref{eq:UPDF}) coincides with Eq. (\ref{eq:UPDF-tree}) for $\mu_Y\sim\mu_R\sim\mu_F\sim t$. Taking the derivative in Eq. (\ref{eq:UPDF-deriv-form}) with the help of the following from of DGLAP equations for PDFs:
\begin{eqnarray}
\frac{d}{d\ln t}\tilde{f}_i(x,t) &=& \frac{\alpha_s(t)}{2\pi} \left[ \sum\limits_j \int\limits_x^1 dz \left[1-\delta_{ij}\theta(z-1+\delta_0)\right] P_{ij}(z) \tilde{f}_j\left( \frac{x}{z},t \right) \right. \nonumber \\
&-& \left. x\tilde{f}_i(x,t) \sum\limits_j \int\limits_0^{1-\delta_0} dz\ zP_{ji}(z) \right], \label{eq:DGLAP}
\end{eqnarray}
which in the limit $\delta_0\to 0$ is exactly equivalent to the usual form of DGLAP equations with (+)-distributions, one obtains:
\begin{eqnarray}
t\Phi_i(x,t,\mu^2)&=& T_i(t,\mu^2,x) \frac{\alpha_s(t)}{2\pi}\int\limits_x^1 dz \left[1-\delta_{ij}\theta(z-1+\delta_0)\right] P_{ij}(x) \tilde{f}_j\left(\frac{x}{z},t \right) \nonumber \\
&+& \tilde{f}_i(x,t) \left[ \frac{d}{d\ln t} T_i(t,\mu^2,x) - T_i(t,\mu^2,x) \frac{\alpha_s(t)}{2\pi} \sum\limits_j\int\limits_0^{1-\delta_0} dz\ zP_{ji}(z) \right] . \label{eq:deriv}
\end{eqnarray}  

  To make contact with Eq. (\ref{eq:UPDF}), one inserts the identity:
\[
  1=\theta(\Delta(t,\mu^2)-z) + \theta(z-\Delta(t,\mu^2)),
\]  
  into the $z$-integrands in Eq. (\ref{eq:deriv}). Then each integral over $z$ can be split in two terms with integrations over regions $x \leq z \leq \Delta (t,\mu^2)$ and $\Delta(t,\mu^2)<z\leq 1-\delta_0$ (assuming that $\Delta(t,\mu^2)<1-\delta_0$) and after reshuffling of some terms, one obtains:
\begin{eqnarray}
&&t\Phi_i(x,t,\mu^2)= T_i(t,\mu^2,x) \frac{\alpha_s(t)}{2\pi}\int\limits_x^1 dz\  P_{ij}(z) \tilde{f}_j\left(\frac{x}{z},t \right) \theta(\Delta(t,\mu^2)-z) \nonumber \\
&&+ \tilde{f}_i(x,t) \left\{ \frac{d}{d\ln t} T_i(t,\mu^2,x) - T_i(t,\mu^2,x) \frac{\alpha_s(t)}{2\pi} \left[ \sum\limits_j\int\limits_0^1 dz\ zP_{ji}(z)\theta(\Delta(t,\mu^2)-z) \right. \right. \nonumber \\
&&+ \left. \left. \int\limits_{\Delta(t,\mu^2)}^1 dz\ \sum\limits_j \left( zP_{ji}(z) \theta(1-\delta_0-z) - \frac{\tilde{f}_j\left(\frac{x}{z},t \right)}{\tilde{f}_i(x,t)} P_{ij}(z) \left[1-\delta_{ij}\theta(z-1+\delta_0)\right]   \right)  \right] \right\} . \label{eq:der-2}
\end{eqnarray}
  In the first line of this equation we have got exactly Eq.(\ref{eq:UPDF}), therefore we have to put to zero the expression in curly brackets in Eq.~(\ref{eq:der-2}), which leads to a differential equation for $T_i(t,\mu^2,x)$. Another important observation is, that one can safely put $\delta_0=0$ in Eq. (\ref{eq:der-2}). Indeed, if $i\neq j$ then there is no singularity in $P_{ij}(z)$ at $z=1$ so integral just converges, while if $i=j$, then singularity at $z=1$ cancels between two terms in the inner-most circular brackets, so integral over $z$ is convergent for $\delta_0=0$ anyway.  

  The solution of obtained differential equation for Sudakov formfactor, satisfying boundary condition $T_i(t=\mu^2,\mu^2,x)=1$ has the form:
\begin{equation}
T_i(t,\mu^2,x)=\exp\left[ -\int\limits_t^{\mu^2} \frac{dt'}{t'} \frac{\alpha_s(t')}{2\pi} \left( \tau_i(t',\mu^2) + \Delta\tau_i (t',\mu^2,x) \right) \right],\label{eq:Sudakov}
\end{equation}
with
\begin{eqnarray}
\tau_i(t,\mu^2)&=&\sum\limits_j \int\limits_0^1 dz\ zP_{ji}(z)\theta (\Delta(t,\mu^2)-z), \label{eq:tau} \\
\Delta\tau_i(t,\mu^2,x)&=& \sum\limits_j \int\limits_0^1 dz\ \theta(z-\Delta(t,\mu^2)) \left[ zP_{ji}(z) - \frac{\tilde{f}_j\left(\frac{x}{z},t \right)}{\tilde{f}_i(x,t)} P_{ij}(z) \theta(z-x) \right]. \label{eq:dtau}
\end{eqnarray}

 We have written these formulas in the Ref.~\cite{NS_DIS1} for the first time, without a detailed derivation. The Sudakov formfactor without the $\Delta \tau_i$-term in the exponent is similar to the Sudakov formfactor of LO KMRW UPDF of Ref.~\cite{Martin:2009ii} but with a numerically-important difference that in our MMRK approach, the rapidity-ordering condition is imposed both on quarks and gluons, while in KMRW-approach it is imposed only on gluons. The term proportional to the ratio of PDFs in Eq. (\ref{eq:dtau}) is familiar from the expression for parton non-emission porbability in ``unitary'' Parton Showers~\cite{Buckley:2011ms}. Strictly-speaking, this term makes transformation from PDF to UPDF non-linear w.r.t. the former. 

 Important property of Eqns.~(\ref{eq:UPDF}), (\ref{eq:tau}), (\ref{eq:dtau}) is that they guarantee exact equivalence of definitions (\ref{eq:UPDF-deriv-form}) and (\ref{eq:UPDF}) at any order in $\alpha_s$ and scheme-choice for DGLAP splitting functions $P_{ij}(z)$ as soon as the PDFs $\tilde{f}_i(x,\mu^2)$ satisfy usual DGLAP equations with the same splitting functions. For alternative ways to ensure the exact normalization condition (\ref{eq:norm-cond}) for KMRW-type UPDF see Ref.~\cite{Guiot:2019vsm}.

\section{Comparison with Collins-Soper-Sterman fromalism} 
\label{sec:PRA-CSS}
For the hadroproduction of Drell-Yan lepton pairs with ${\bf q}_T^2\ll Q^2$ the perturbative resummation of higher-order corrections enhanced by $\ln(Q^2/{\bf q}_T^2)$ is performed by Collins-Soper-Sterman formula~\cite{Collins:1984kg}:
 \begin{eqnarray}
 \frac{d\sigma}{dQ^2 d{\bf q}_T^2 dy} &=& \frac{\alpha}{3\pi Q^2 Q_T^4} \sum\limits_{j,a,b} \frac{(4\pi \alpha) e_j^2}{4N_c} \int d^2{\bf x}_T e^{i{\bf q}_T {\bf x}_T} \left[ \int\limits_{x_+}^1 \frac{dz_+}{z_+} \tilde{f}_a\left(\frac{x_+}{z_+},\frac{1}{{\bf x}_T^2}\right) C_{ja}(z_+ ,\alpha_s(1/{\bf x}_T^2) ) \right] \nonumber \\ 
&\times& \left[ \int\limits_{x_-}^1 \frac{dz_-}{z_-} \tilde{f}_a\left(\frac{x_-}{z_-},\frac{1}{{\bf x}_T^2}\right) C_{ja}(z_- ,\alpha_s(1/{\bf x}_T^2) ) \right]\times S({\bf x}_T^2,Q^2),  
 \end{eqnarray}
where ${\bf x}_T$ is a transverse coordinate, conjugated to transverse-momentum ${\bf q}_T$, $C_{ij}$ are the collinear matching-functions, which are usually taken order-by-order in $\alpha_s$ and the resummation is performed by Sudakov formfactor in the ${\bf x}_T$-space:
\begin{equation}
S({\bf x}_T^2,Q^2)=\exp\left[ -\int\limits_{1/{\bf x}_T^2}^{Q^2} \frac{dt'}{t'}\left( A(\alpha_s(t'))\ln\frac{Q^2}{t'} + B(\alpha_s(t')) \right) \right],\label{eq:Sud-CSS}
\end{equation}
where functions $A$ and $B$, corresponding respectively to the resummation of doubly ($\propto \ln^2 ({\bf x}_T^2 Q^2)$) and single-logarithmic ($\propto \ln ({\bf x}_T^2Q^2)$) corrections admit the following perturbative expansion (Eqns. (3.18) and (3.20) in~\cite{Collins:1984kg}):
\begin{eqnarray}
A(\alpha_s)&=&C_F\frac{\alpha_s}{\pi} + O(\alpha_s^2), \label{eq:A-CSS} \\
B(\alpha_s)&=&2C_F\left[ -\frac{3}{4} + \ln \frac{C_1}{2C_2} + \gamma_E \right] \frac{\alpha_s}{\pi} + O(\alpha_s^2), \label{eq:B-CSS}
\end{eqnarray}
where we have explicitly shown terms up to Next-to-Leading Logarithmic (NLL) Approximation. The NLL coefficient $B$ in Eq. (\ref{eq:B-CSS}) depends on the resummation scheme, which is defined by parameters $C_{1,2}$ in the Ref.~\cite{Collins:1984kg}.  

  On our momentum-space language, the formfactor (\ref{eq:Sud-CSS}) corresponds to the convolution of two UPDFs in transverse-momentum space, so one should compare the logarithmic structure of our formfactor (\ref{eq:Sudakov}) with a square-root of the formfactor (\ref{eq:Sud-CSS}). Substituting the leading-order expressions for DGLAP splitting functions to Eq. (\ref{eq:tau}), 
one obtains:
\[
\tau_q(\Delta) = \int\limits_0^{\Delta}dz\ z\left( P_{qq}(z) + P_{gq}(z) \right) = C_F \left[ -2\ln(1-\Delta) - \frac{3}{2} \right] + O(1-\Delta).
\]
  The correction $\Delta \tau_q$ is a quantity $O(1-\Delta)$, so it contributes only beyond NLL-approximation.  Substituting the last result for $\tau_q$ into Eq.(\ref{eq:Sudakov}) and taking into account, that for $t\ll \mu^2$: $1-\Delta(t,\mu^2)\simeq \sqrt{t/\mu^2}$, one obtains in this limit:
\begin{equation}
T_q(t,\mu^2) \simeq \exp\left[ -\frac{\alpha_s}{2\pi} C_F \left( \frac{1}{2}\ln^2 \frac{\mu^2}{t} - \frac{3}{2}\ln \frac{\mu^2}{t} \right)  \right], \label{eq:Sudakov-NLL}
\end{equation}  
 where we have taken into account, that running-coupling effects in Eq.(\ref{eq:Sudakov}) also contribute only beyond NLL as well as effects of scale-dependence of the PDF in Eq.~ Eq.(\ref{eq:UPDF-deriv-form}). So one should consider only the Fourier-transform of a derivative $dT_{qq}({\bf q}_T^2,\mu^2)/d{\bf q}_T^2$.  Taking the Fourier-transform of a $t$-derivative of Eq. (\ref{eq:Sudakov-NLL}) order-by-order in $\alpha_s$, with the help of the relation:
\[
\frac{1}{{\bf q}_T^2} \ln^n \frac{\mu^2}{{\bf q}_T^2} \to \frac{-1}{n+1}\ln^{n+1} (\mu^2 {\bf x}_T^2)+\ldots,
\]
where by ellipsis we denote non-logarithmic terms, one obtains:
\[
\frac{dT_q({\bf q}_T^2,\mu^2)}{d{\bf q}_T^2} \to \exp\left[ -\frac{\alpha_s}{2\pi} C_F \left( \frac{1}{2}\ln^2(\mu^2{\bf x}_T^2) - \frac{3}{2} \ln (\mu^2{\bf x}_T^2) \right) \right].
\]

The last result indeed coincides with the square-root of Eq.~(\ref{eq:Sud-CSS}) with coefficients (\ref{eq:A-CSS}) and (\ref{eq:B-CSS}) taken up to NLL-approximation in a scheme with $\ln C_1/(2C_2)=-\gamma_E$. So we conclude, that our resummation scheme is consistent with perturbative part of CSS formalism up to NLL-approximation in the region ${\bf q}_T^2\ll Q^2$ where both formalisms apply, thanks to a particular small-$t$ asymptotics of the KMRW cutoff function: $1-\Delta(t,\mu^2)\simeq \sqrt{t/\mu^2}$.   

\section{Drell-Yan lepton pair production in PRA}
\label{sec:PRA-DY}
 The LO in $\alpha_s$ cross-section of $p(P_1)+p(P_2)\to l^+(p_1) + l^-(p_2) + X$-process in PRA is given by:
\begin{eqnarray}
\frac{d\sigma}{dQ^2} &=& \sum\limits_{i,j}\int\limits_0^1\frac{dx_1}{x_1} \int\frac{d^2{\bf q}_{T1}}{\pi} \Phi_i(x_1,{\bf q}_{T1}^2,\mu_Y^2) \int\limits_0^1\frac{dx_2}{x_2} \int\frac{d^2{\bf q}_{T2}}{\pi} \Phi_j(x_2,{\bf q}_{T2}^2,\mu_Y^2) \nonumber \\
&\times & \int \frac{d^4 q}{(2\pi)^3} \delta(q^2-Q^2) \frac{(2\pi)^4}{2Sx_1x_2} \delta^{(4)}(q_1+q_2-q) \nonumber \\
&\times & (2\pi)^3\int \frac{d^4 p_1 d^4 p_2}{(2\pi)^6} \delta_+(p_1^2-m_l^2) \delta_+(p_2^2-m_l^2) \delta^{(4)}(q-p_1-p_2)  \overline{|{\cal M}_{ij}|^2}, \label{eq:CS_2-2_PRA}
\end{eqnarray}
where we have introduced an integration over intermediate momentum $q=p_1+p_2$, parton momenta are given by $q_1^\mu=x_1P_1^\mu+q_{T1}^\mu$ and $q_2^\mu=x_2P_2^\mu+{q}_{T2}^\mu$ and PRA squared matrix element $\overline{|{\cal M}_{ij}|^2}$ is a function of scalar products of four-momenta of partons($q_{1,2}$), leptons($p_{1,2}$) and vectors $n_{+}$ or $n_{-}$.

 In the last line of Eq.~(\ref{eq:CS_2-2_PRA}) one can integrate-out $p_2$ using the delta-function and then pass to the center-of-mass frame of the lepton pair, to express this integral in terms of spherical angles $\theta_{l^+}$ and $\phi_{l^+}$, parametrizing the direction of lepton momentum in this frame. 

  In first two lines of Eq.(\ref{eq:CS_2-2_PRA}) one integrates-out momentum $q$ and momentum-fraction $x_2$, using the relation:
\[
\delta\left( (q_1+q_2)^2-Q^2 \right) = \frac{1}{Sx_1}\delta\left( x_2-\frac{Q_T^2}{Sx_1} \right),
\] 
  and replaces $d^2{\bf q}_{T2}\to \pi d{\bf q}_T^2$ to finally obtain:
\begin{equation}
 \frac{d\sigma}{dQ^2 d{\bf q}_T^2 dy d\Omega_l} = \int\limits_0^\infty \frac{d{\bf q}_{T1}^2}{2} \int\limits_0^{2\pi} d\phi_1\ \Phi_i(x_1,{\bf q}_{T1}^2,\mu_Y^2) \Phi_j(x_2,{\bf q}_{T2}^2,\mu_Y^2) \frac{\sqrt{Q^2-4m_l^2}}{4Q(2\pi)^3} \frac{\overline{|{\cal M}_{ij}|^2}}{Q_T^4},\label{eq:CS_2-2_PRA-MASTER}
\end{equation}
where $y$ is the rapidity of the lepton pair, so that $x_{1,2}=Q_Te^{\pm y}/\sqrt{S}$ and $\phi_1$ is the azimuthal angle of ${\bf q}_{T1}$, while ${\bf q}_{T2}={\bf q}_T-{\bf q}_{T1}$.  If four-momenta $q^\mu=(Q_T{\rm ch}y,|{\bf q}_T|,0,Q_T{\rm sh}y)^\mu$ and $q_{1,2}^\mu=(\sqrt{S}x_{1,2}/2,{\bf q}_{T1,2},\pm\sqrt{S}x_{1,2}/2)^\mu$ are given in the $pp$ center-of-mass frame, then four-momenta of leptons can be expressed using covariant relations:
\[
p_{1,2}^\mu=\frac{q^\mu}{2}\pm \frac{\sqrt{Q^2-4m_l^2}}{2}\left[ X^\mu \sin\theta_l\cos\phi_l + Y^\mu\sin\theta_l\sin\phi_l + Z^\mu\cos\theta_l \right],
\] 
with the help of following expressions for $pp$ center-of-mass frame components of unit vectors $X^\mu$, $Y^\mu$ and $Z^\mu$ of the Collins-Soper frame~\cite{Collins:1977iv}:
\begin{eqnarray*}
X^\mu&=&\left( \frac{|{\bf q}_T|}{Q}{\rm ch}y, \frac{Q_T}{Q},0,\frac{|{\bf q}_T|}{Q}{\rm sh}y  \right)^\mu,\\
Y^\mu&=&\mathop{\rm sgn} y\times\left( 0,0,1,0 \right)^\mu, \\
Z^\mu&=&\mathop{\rm sgn} y\times\left( {\rm sh}y, 0,0,{\rm ch}y  \right)^\mu,
\end{eqnarray*}
where factors $\mathop{\rm sgn} y$ take into account that the direction of $z$-axis of Collins-Soper frame  in the analysis of experimental data in Ref.~\cite{ATLAS-2016} coincides with the positive direction of longitudinal projection of a vector boson momentum in the $pp$ center of mass frame.

 Explicit expressions for components of all vectors given in one reference frame allow us to calculate all scalar products which $\overline{|{\cal M}_{ij}|^2}$ depends upon and set up experimental cuts on momenta of leptons $p_{1,2}$. 

 The squared PRA amplitude of the LO in $\alpha_s$ partonic subprocess:
\begin{equation}
Q(q_1)+\bar{Q}(q_2)\to \gamma^*/Z^* \to l^+(p_1) + l^-(p_2),\label{proc:QQ-ll}
\end{equation}
where by $Q$($\bar{Q}$) we denote Reggeized quark(anti-quark) is given by:
\begin{eqnarray}
\overline{|{\cal M}_{q\bar{q}}|^2}&=&(4\pi\alpha)\left\{ \frac{1}{\hat{s}^2} w^{(q\bar{q},VV,\gamma\gamma)}_{\mu\mu} L^{\mu\nu}_{VV} \right. +\frac{1}{(\hat{s}-M_Z^2)^2+M_Z^2\Gamma_Z^2}\sum_{a,b,\bar{a},\bar{b}=V,A} C^{(b)}_{Zl^-} C^{(\bar{b})}_{Zl^+} w^{(q\bar{q},a\bar{a},ZZ)}_{\mu\mu} L^{\mu\nu}_{b\bar{b}} \nonumber  \\
&+&\left. \frac{4(\hat{s}-M_Z^2)}{\hat{s}\left( (\hat{s}-M_Z^2)^2+M_Z^2\Gamma_Z^2 \right)} \sum_{a,b=V,A} C^{(b)}_{Zl^-} w^{(q\bar{q},aV,Z\gamma)}_{\mu\mu} L^{\mu\nu}_{bV} \right\}, \label{eq:M2_QQ-ll_gen}
\end{eqnarray}
where the first, second and third terms in curly brackets correcpond to the squared photon, $Z$-boson exchange diagrams and $Z^*\gamma^*$-interference respectively.  In our numerical calculations we have taken $m_l=0$, however here we write-down formulas for $m_l\neq 0$ for generality. The leptonic tensor in Eq.~(\ref{eq:M2_QQ-ll_gen}) is given by standard expression:
\[
L^{\mu\nu}_{b\bar{b}}={\rm tr}\left[ \left( \hat{p}_1+m_l \right)\gamma^\mu (\delta_{bV}+\delta_{bA}\gamma_5) \left( \hat{p}_2-m_l \right) (\delta_{\bar{b}V}-\delta_{\bar{b}A}\gamma_5) \gamma^\nu   \right],
\]
while PRA partonic tensor $w^{(q\bar{q},a\bar{a})}_{\mu\nu}$ is given in Eq.~(\ref{eq:w-PRA}).

  After taking all traces and index-contractions, the squared amplitude can be simplified as follows:   
\begin{eqnarray}
\overline{|{\cal M}_{q\bar{q}}|^2}&=&\frac{(4\pi\alpha)^2}{4N_c}\left\{ \frac{e_q^2}{\hat{s}^2}M^2_{VV,VV} + \frac{1}{(\hat{s}-M_Z^2)^2+M_Z^2\Gamma_Z^2} \left[ 2\left(C^{(q)}_{AA}C^{(q)}_{VV} + C^{(q)}_{AV}C^{(q)}_{VA}\right)M^2_{VV,AA} \right. \right. \nonumber \\
&+& \left. \left( (C^{(q)}_{AA})^2 + (C^{(q)}_{AV})^2 +(C^{(q)}_{VA})^2 + (C^{(q)}_{VV})^2  \right) M^2_{VV,VV} + \left( (C^{(q)}_{AA})^2 + (C^{(q)}_{VA})^2 \right) \Delta M^2 \right] \nonumber \\
&+& \left. \frac{4(\hat{s}-M_Z^2) e_q}{\hat{s}\left( (\hat{s}-M_Z^2)^2+M_Z^2\Gamma_Z^2 \right)} \left[ C_{VV}^{(q)} M^2_{VV,VV} + C_{AA}^{(q)} M^2_{VV,AA} \right] \right\},\label{eq:M2-PRA-final}
\end{eqnarray}
where $C_{ab}^{(q)}=C_q^{(a)}C_{Zl^-}^{(b)}$ is the product of quark and lepton coupling-factors (\ref{eq:C-fact}), while
\begin{eqnarray*}
M_{VV,VV}^2&=& \frac{8}{q_1^+ q_2^-} \left[ 2 (p_1^-  q_1^+)^2  (q_1^+  q_2^- - t_1) \right. \\
&-& 2  p_1^-  q_1^+  q_2^-  
 \left( q_1^+  (m_l^2 - \hat{t} - t_1) + 
  p_1^+  (2  q_1^+  q_2^- + \hat{t} + \hat{u} -2  m_l^2) \right) \\
&+& q_2^-  \left(2  (p_1^+)^2  q_2^-  (q_1^+  q_2^- - t_2) + 
  2  p_1^+  q_1^+  q_2^-  (t_2 + \hat{u} -m_l^2) \right. \\
&+& 
  \left. \left. q_1^+  (2  m_l^2  (q_1^+  q_2^- + \hat{s}) - \hat{s}  (\hat{t} + \hat{u})) \right) \right], \\
M_{VV,AA}^2&=& 8\hat{s}\left( \hat{u}-\hat{t}+2(p_1^+q_2^- - p_1^-q_1^+) \right), \\
\Delta M^2 &=& 32 m_l^2 \left( \hat{t}+\hat{u}-2m_l^2 \right), 
\end{eqnarray*}
where $\hat{s}=(q_1+q_2)^2$, $\hat{t}=(q_1-p_1)^2$, $\hat{u}=(q_2-p_1)^2$ and $t_{1,2}={\bf q}_{T1,2}^2$. If instead of the process (\ref{proc:QQ-ll}) one considers the process $\bar{Q}(q_1)+Q(q_2)\to l^+ + l^-$, the overall sign of the Lorentz-structure $M_{VV,AA}^2$ should be flipped.

\section{Non-perturbative part of the UPDF, fit to low-energy data}
\label{sec:UPDF-fit}
Perturbative expression (\ref{eq:UPDF}) does not define the UPDF for all values of $t$, since for $t<\Lambda_{\rm QCD}^2$, the integral in Eq. (\ref{eq:Sudakov}) is ill-defined due to Landau pole of $\alpha_s(t')$. Similar problem arises also in TMD-factorization~\cite{CollinsQCD} in a form of non-perturbative ambiguity of the rapidity-evolution kernel~\cite{Scimemi:2019cmh}. Analogously with the Ref.~\cite{Martin:2009ii} we define the UPDF for $t<t_0=1$ GeV as:
\[
t\Phi_i(x,t,\mu^2)=At^{1+\alpha}(t_1-t),
\]  
where parameters $A$, $t_1$ and $\alpha$ are determined from the requirements of normalisation of UPDF (\ref{eq:norm-cond}), it's continuity, smoothness in the point $t=t_0$ and positivity of $\alpha$ as follows:
\begin{eqnarray*}
A&=&\frac{F_2t_0^{-1-\alpha}}{t_1-t_0},\ t_1=\frac{t_0 (1+\alpha) (2+\alpha-\beta) }{(2+\alpha)(1+\alpha-\beta)}, \\
\alpha&=&\max\left[\beta-\frac{3}{2}+\frac{1}{2}\sqrt{1+4\beta(\beta+\beta_1)}, 0\right],
\end{eqnarray*}
where $\beta=F_2/F_1$, $\beta_1=(F_3-F_2)/t_0/F_2/\delta$, $F_1=\tilde{f}_i(x,t_0)T(t_0,\mu^2,x)$, $F_2=t\Phi_i^{\rm (pert.)}(x,t_0,\mu^2)$, $F_3=t\Phi_i^{\rm (pert.)}(x,t_0(1+\delta),\mu^2)$ and $\delta\ll 1$. 

 The UPDF defined for all values of $t$ as described above we call the {\it shower-part} of the UPDF. Physically it is determined by perturbative dynamics of QCD for $t>t_0$ and non-perturbative properties of QCD vacuum for $t<t_0$~\cite{Vladimirov:2020umg, Vladimirov:2020ofp}. To take into account non-perturbative intrinsic motion of partons inside hadron we convolute the shower part on UPDF with phenomenological intrinsic transverse-momentum distribution, which we take in the $x$-independent Gaussian from:
\begin{equation}
\Phi_i(x,{\bf q}_T^2,\mu^2)= \int\frac{d^2{\bf k}_T}{\pi \sigma_{Ti}^2} e^{-\frac{{\bf k}_T^2}{\sigma_{Ti}^2}}  \Phi^{\rm (shower)}_i(x,({\bf q}_T-{\bf k}_T)^2,\mu^2).
\end{equation}
The non-perturbative parameters $\sigma_{Ti}$ will be determined below from a fit of experimental data with fine binning in $|{\bf q}_T|<1$ GeV region, but from physical interpretation given above one expects to find $\sigma_{Ti}\sim\Lambda_{QCD}\ll 1$ GeV. 

  We perform the global fit of parameters $\sigma_{Ti}$, using experimental data on Drell-Yan lepton pair production at $\sqrt{S}\leq 200$ GeV, summarized in the Tab.~\ref{tab:data}. To obtain the shower part of the UPDF with the procedure described above, we use the LO PDF set MSTW-2008~\cite{Martin:2009iq} and formulas~(\ref{eq:UPDF}), (\ref{eq:Sudakov}) -- (\ref{eq:dtau}) with LO DGLAP splitting functions substituted. We also adopt the scale-choice $\mu_Y=\xi Q_T$ in our numerical calculations, with the default value for $\xi=1$ for the central curves and $\xi=2^{\pm 1}$ for the boundaries of scale-uncertainty bands which are shown as gray corridors in the figures below.

  Since we do not expect our formalism to describe overall normalization of the data, due to the lack of complete NLO loop corrections, we perform the comparison with normalized distributions $(1/\sigma)d\sigma/d{\bf q}_T^2$. To this end we multiply each experimental spectrum by constant, ${\bf q}_T-$independent factor, obtained via the summation of central-values of cross-section in all data-bins. In the Tab.~\ref{tab:data} we show the obtained ratios of experimental total cross-sections and our theoretical results. We also present the uncertainties on this $K$-factors due to the scale-variation in theoretical predictions (with experimental cross-sections fixed at their central values) and due to experimental uncertainties (divided by central theoretical predictions). 

\begin{table}
\begin{tabular}{|c|c|c|c|c|}
\hline
Dataset & Observable & $\sqrt{S}$(GeV) & $Q$(GeV)  & $\frac{\sigma({\rm data})}{\sigma{\rm (theory)}}$ [+/- scale-uncert.] (+/- exp. uncert.) \\
\hline
\hline
\multirow{14}{*}{E-288~\cite{E-288}} &\multirow{14}{*}{$q^0{d\sigma}/{d^3q}$} & \multirow{5}{*}{19.4} & 4-5 & 1.54[+0.63/-0.40]($\pm 0.20$)  \\
                        &                                           &                       & 5-6 & 1.50[+0.70/-0.45]($\pm 0.18$) \\ 
                        &                                           &                       & 6-7 & 1.43[+0.73/-0.46]($\pm 0.18$)  \\
                        &                                           &                       & 7-8 & 1.22[+0.70/-0.43]($\pm 0.25$)  \\
                        &                                           &                       & 8-9 & 1.03[+0.64/-0.04]($\pm 0.35$)  \\ \cline{3-5}
                        &                                           & \multirow{4}{*}{23.7} & 4-5 & 1.64[+0.56/-0.35]($\pm 0.22$)  \\
                        &                                           &                       & 5-6 & 1.46[+0.57/-0.36]($\pm 0.14$) \\
                        &                                           &                       & 6-7 & 1.47[+0.64/-0.42]($\pm 0.17$)  \\
                        &                                           &                       & 7-8 & 1.47[+0.70/-0.44]($\pm 0.20$)  \\
                        &                                           &                       & 8-9 & 1.43[+0.71/-0.45]($\pm 0.29$)  \\ \cline{3-5}
                        &                                           & \multirow{4}{*}{27.4} & 5-6 & 1.57[+0.55/-0.33]($\pm 0.13$)  \\
                        &                                           &                       & 6-7 & 1.47[+0.57/-0.36]($\pm 0.07$)  \\
                        &                                           &                       & 7-8 & 1.44[+0.60/-0.38]($\pm 0.08$)  \\
                        &                                           &                       & 8-9 & 1.35[+0.60/-0.38]($\pm 0.10$)  \\ \hline 
\multirow{5}{*}{E-605~\cite{E-605}}  &\multirow{5}{*}{$q^0{d\sigma}/{d^3q}$}& \multirow{5}{*}{38.8} & 7-8 &  1.50[+0.55/-0.31]($\pm 0.18$) \\
                        &                                           &                       & 8-9 & 1.42[+0.56/-0.33]($\pm 0.10$)  \\
                        &                                           &                       & 10.5-11.5 & 1.33[+0.60/-0.38]($\pm 0.11$) \\
                        &                                           &                       & 11.5-13.5 & 1.40[+0.67/-0.40]($\pm 0.11$) \\
                        &                                           &                       & 13.5-18   & 1.14[+0.60/-0.36]($\pm 0.17$) \\ \hline
                R-209~\cite{R-209}   & ${d\sigma}/{d{\bf q}_T^2}$            &                62     & 5-8       & 1.63[+0.40/-0.18]($\pm 0.29$) \\ \hline
               PHENIX~\cite{PHENIX}   & $q^0{d\sigma}/{d^3q}$                 &               200     & 4.8-8.2   & 1.50[+0.17/-0.10]($\pm 0.44$) \\ \hline\hline
CDF-1999~\cite{CDF-1999}& $d\sigma/d|{\bf q}_T|$ & 1800 & 66-116 & 2.07 [+0.23/-0.12]($\pm 0.11$) \\ \hline
ATLAS-2019~\cite{ATLAS-2019}& $d\sigma/d|{\bf q}_T|$ & 13000 & 66-116 & 1.71 [+0.07/-0.06]($\pm 0.04$) \\ \hline                 
\end{tabular}
\caption{Experimental data on transverse-momentum spectra of Drell-Yan lepton pairs used in the present study and corresponding values of $K$-factors. The ATLAS-2019 and CDF-1999 data are not included into the fit of $\sigma_{T}$. \label{tab:data}}
\end{table}

  Although in principle, parameters $\sigma_{Ti}$ could be flavor-dependent, due to a large theoretical uncertainty of our LO analysis we have not found any significant improvement in the fit quality from taking different values of $\sigma_{Ti}$ for different falvors or for sea vs. valence quarks. Therefore we present only the results with all $\sigma_{Ti}$ taken to be equal to the same constant $\sigma_T$, for which we have found $\sigma_T^{\rm (best\ fit)}=0.35$ GeV, leading to the $\chi^2/$d.o.f$=1.5$ with total of 323 data-points in our data-set.

\begin{figure}
\begin{center}
\includegraphics[width=0.7\textwidth]{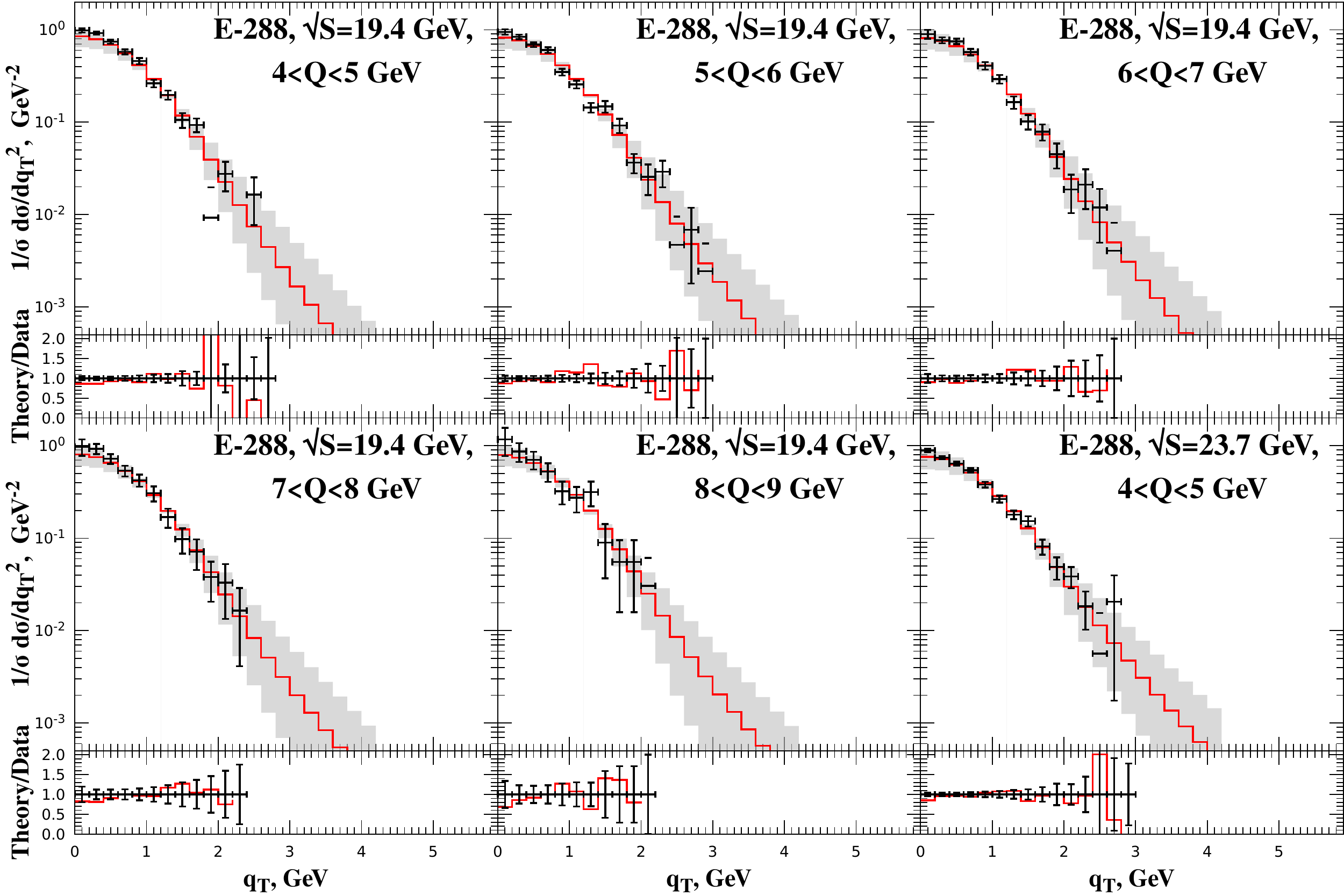}
\end{center}
\caption{Description of normalized $(1/\sigma)d\sigma/d{\bf q}_T^2$-distributions for $\sqrt{S}\leq 200$ GeV with $\sigma_T^{\rm (best\ fit)}=0.35$ GeV \label{fig:fit}}
\end{figure}

\begin{figure}
\begin{center}
\includegraphics[width=0.7\textwidth]{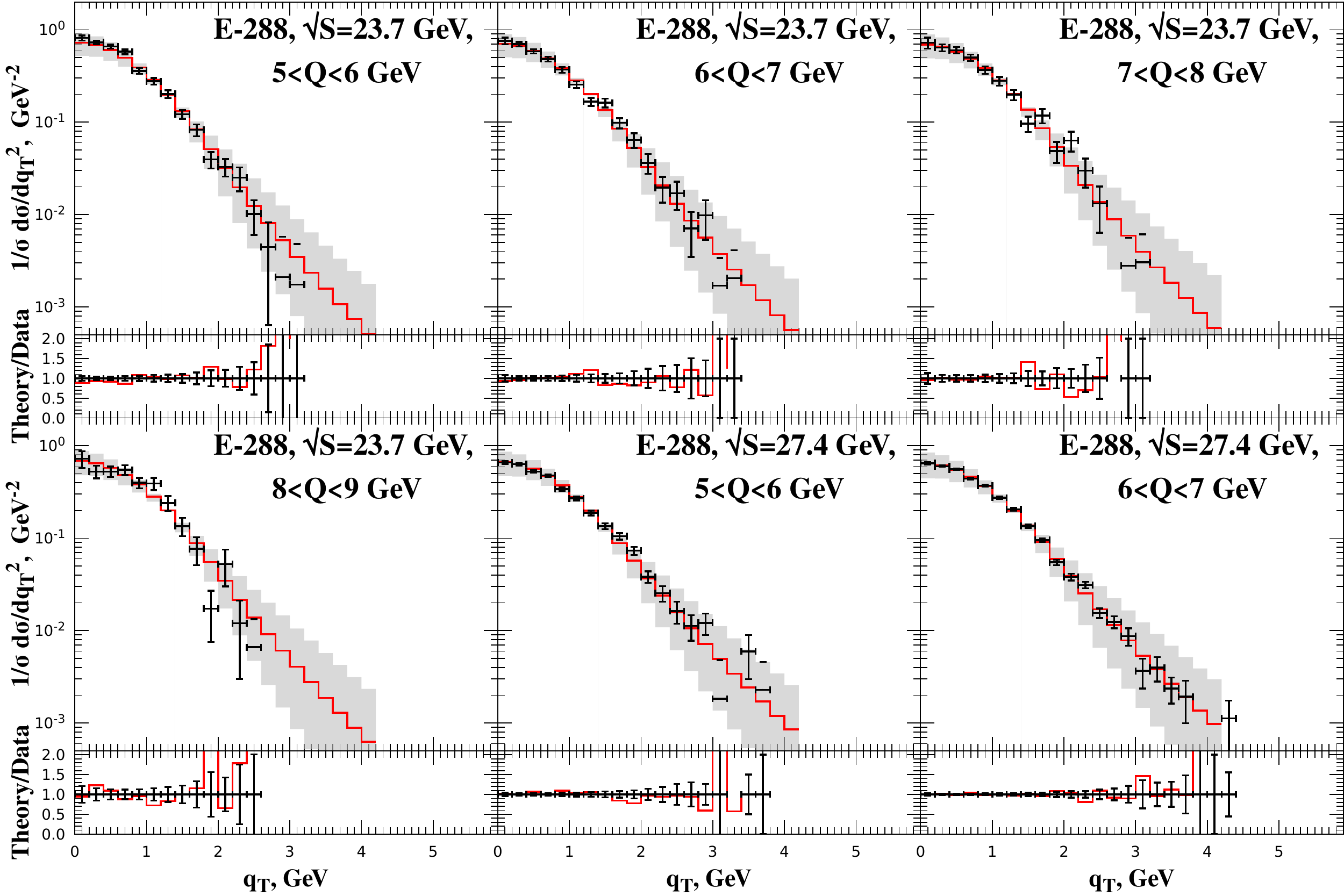}
\end{center}
\caption{Description of normalized $(1/\sigma)d\sigma/d{\bf q}_T^2$-distributions for $\sqrt{S}\leq 200$ GeV with $\sigma_T^{\rm (best\ fit)}=0.35$ GeV. Continuation of the Fig.~\ref{fig:fit} \label{fig:fit1}}
\end{figure}

\begin{figure}
\begin{center}
\includegraphics[width=0.7\textwidth]{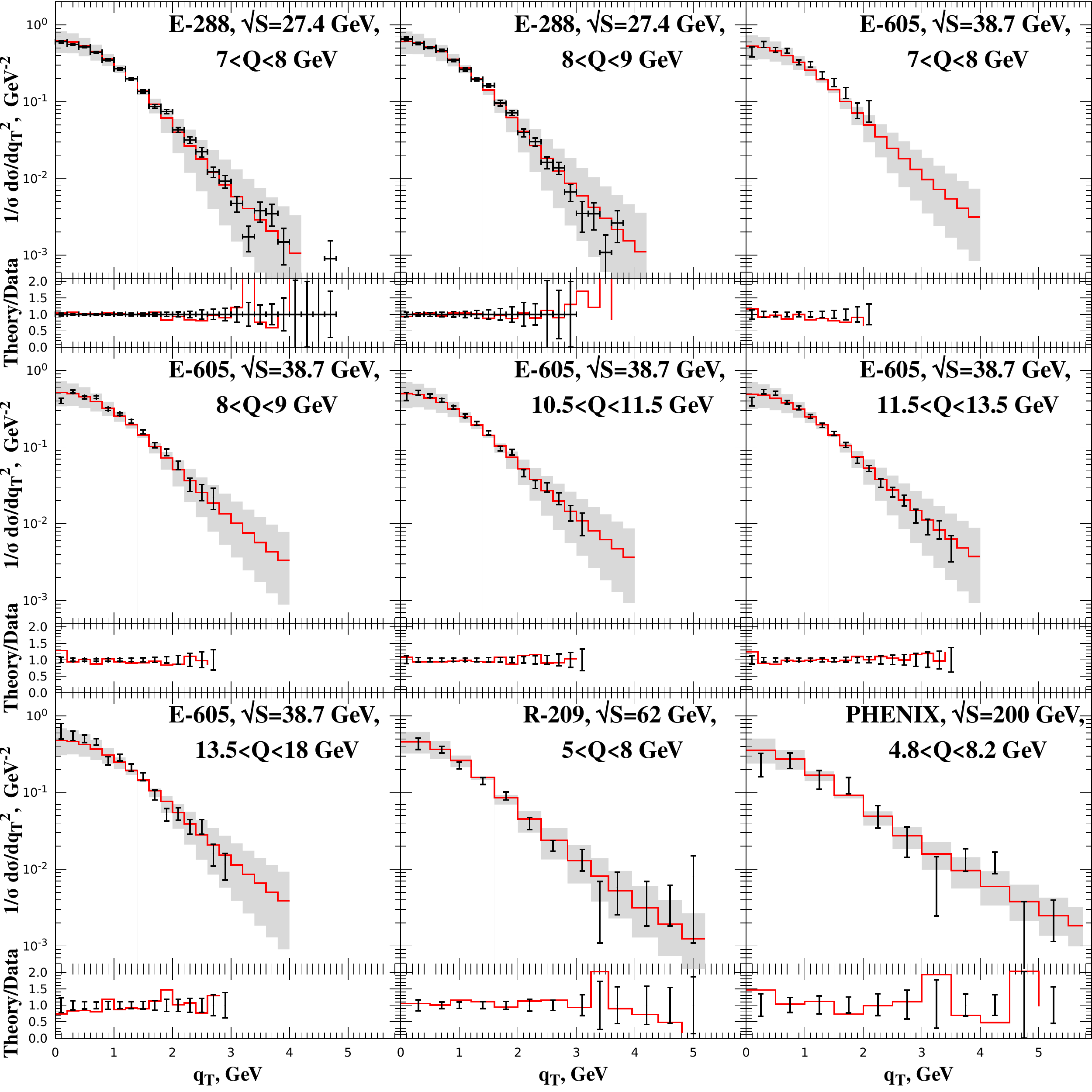}
\end{center}
\caption{Description of normalized $(1/\sigma)d\sigma/d{\bf q}_T^2$-distributions for $\sqrt{S}\leq 200$ GeV with $\sigma_T^{\rm (best\ fit)}=0.35$ GeV. Continuation of Figs.~\ref{fig:fit} and~\ref{fig:fit1} \label{fig:fit2}}
\end{figure}

 The quality of the description of data with this parameters is illustrated in the Figs.~\ref{fig:fit}--\ref{fig:fit2}. From ratio plots provided in this figure one can see, that our central LO PRA prediction describes the shape of ${\bf q}_T$-spectrum for all values of $|{\bf q}_T|$, including the region $|{\bf q}_T|\sim Q$ within experimental uncertainties. As for overall normalisation of the cross-section, one can see form Tab.~\ref{tab:data}, that an overall $K$-factor $\simeq 1.4$ is required to describe the data, which is typical for LO calculations even in the CPM and was also observed in our previous work~\cite{Nefedov:2012cq}.      

\section{Drell-Yan lepton pair production at the Tevatron and LHC}
\label{sec:DY-Tev-LHC}
  In this section we will discuss our predictions for transverse-momentum spectra at the Tevatron and LHC and for ${\bf q}_T$-dependence of angular coefficients, describing polarization of an intermediate vector boson at the LHC energies. To make our predictions we employ the LO UPDF described in Sec.~\ref{sec:UPDF} and \ref{sec:UPDF-fit} as well as NLO UPDF. The shower-part of the latter was generated from the collinear NLO PDFs CTEQ-18~\cite{Dulat:2015mca} using Eqns.~(\ref{eq:UPDF}), (\ref{eq:Sudakov}), (\ref{eq:tau}) and (\ref{eq:dtau}) with the well-known NLO expressions for DGLAP splitting-functions~\cite{Curci:1980uw,Furmanski:1980cm,ellis_stirling_webber_1996} substituted. The non-perturbative part of our NLO UPDF was determined in exactly the same way as for the LO case, as described in Sec.~\ref{sec:UPDF-fit} with the same value of non-perturbative parameter $\sigma_T=0.35$ GeV. Since high-energy data typically have very coarse binning at $|{\bf q}_T|<1$ GeV, the non-perturbative part of the UPDF have negligible effect on our predictions presented in this section. Of course, the usage of NLO UPDF without complete NLO corrections to the PRA hard-scattering coefficient (\ref{eq:M2-PRA-final}) is not fully consistent, however we expect that at least for $|{\bf q}_T|\lesssim Q$ the effect of NLO corrections in PRA on the shape of the distribution will be negligible and NLO correction will affect mostly the overall normalization of the spectrum.

\begin{figure}
\begin{center}
\includegraphics[width=0.5\textwidth]{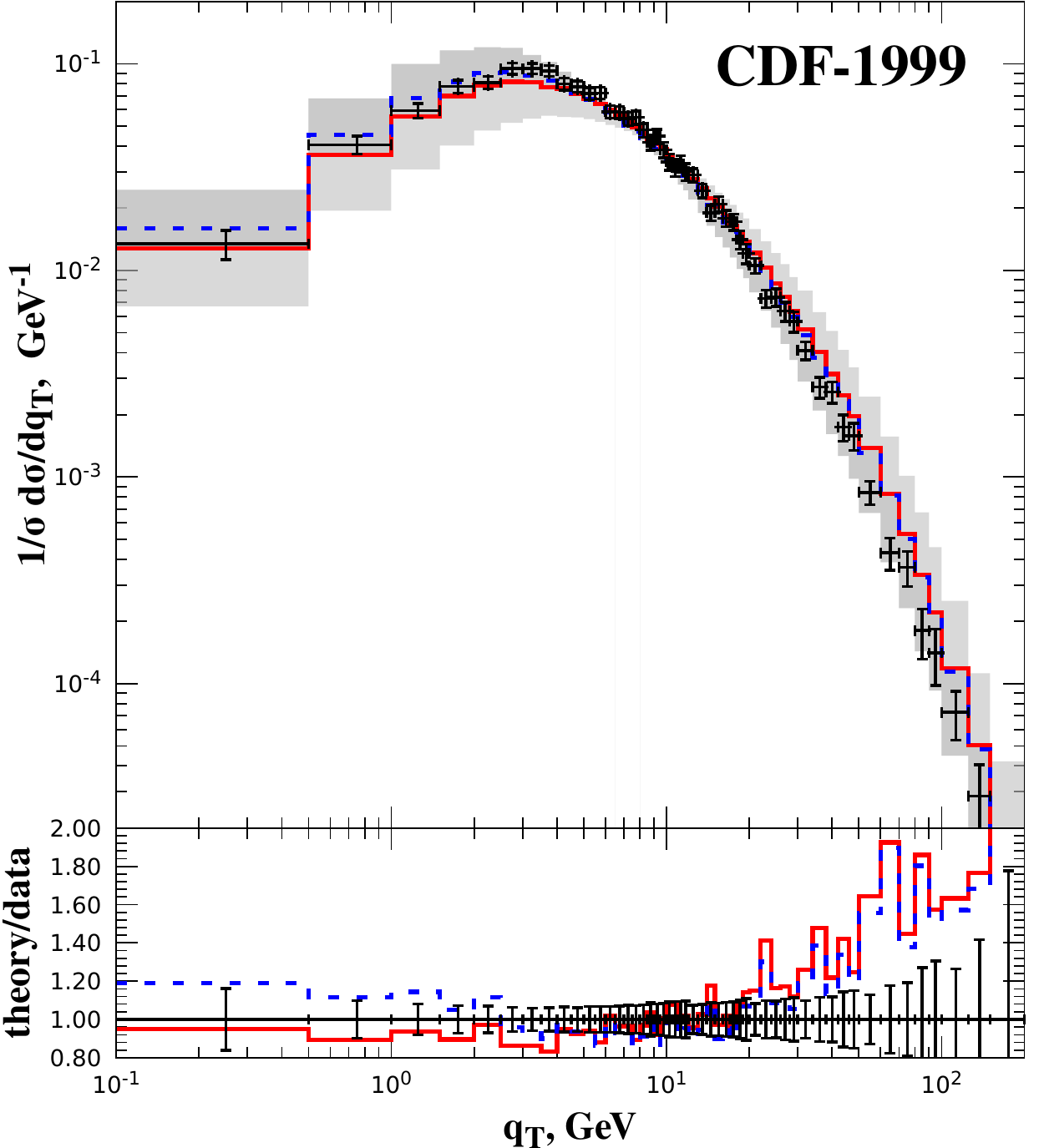}
\end{center}
\caption{The normalized transverse-momentum spectrum of Drell-Yan lepton pairs measured by the CDF collaboration~\cite{CDF-1999} compared to LO PRA predictions made with LO (solid histogram) and NLO (dashed histogram) UPDFs. The uncertainty band is shown only for the LO prediction. \label{fig:Z-CDF}}
\end{figure}

  In the Fig.~\ref{fig:Z-CDF} we compare our predictions for normalized distribution $(1/\sigma)d\sigma/d|{\bf q}_T|$ with experimental data, obtained by CDF collaboration at the Fermilab Tevatron in $p\bar{p}$-collisions with $\sqrt{S}=1.8$ TeV~\cite{CDF-1999}. The measurement has been performed in the dilepton invariant-mass window around $Z$-boson resonance, see Tab.~\ref{tab:data}. A lepton in the rapidity-range $|y_l|<1.1$ had been required to have $p_T^{(l)}>20$ GeV, while for lepton rapidities in the range $1.1<|y_l|<4.2$ a lower cut $p_T^{(l)}>15$ GeV had been imposed in the analysis of Ref.~\cite{CDF-1999}.  

  For the total cross-section in the case of $p\bar{p}$-collisions at higher energies we find a significantly larger $K-$factor $\simeq 2$ as opposed to the typical $K-$factor of $\simeq 1.4$ which we need to describe total cross-section of Drell-Yan process in $p-$nucleon collisions at low energies, see Tab.~\ref{tab:data}. However, our central predictions both with LO and NLO UPDFs describe the shape of $|{\bf q}_T|$-distribution in the Fig.~\ref{fig:Z-CDF} remarkably well, with NLO UPDF providing somewhat better result. The shape of  of $d\sigma/d|{\bf q}_T|/\sigma$-distribution is described within experimental uncertainties for all values of $|{\bf q}_T|/M_Z$ up to $\simeq 0.4$, which corresponds to $|{\bf q}_T|/\sqrt{S}\simeq 0.02$. At higher transverse-momenta, our predictions deviate from experimental data.   

\begin{figure}
\begin{center}
\includegraphics[width=0.49\textwidth]{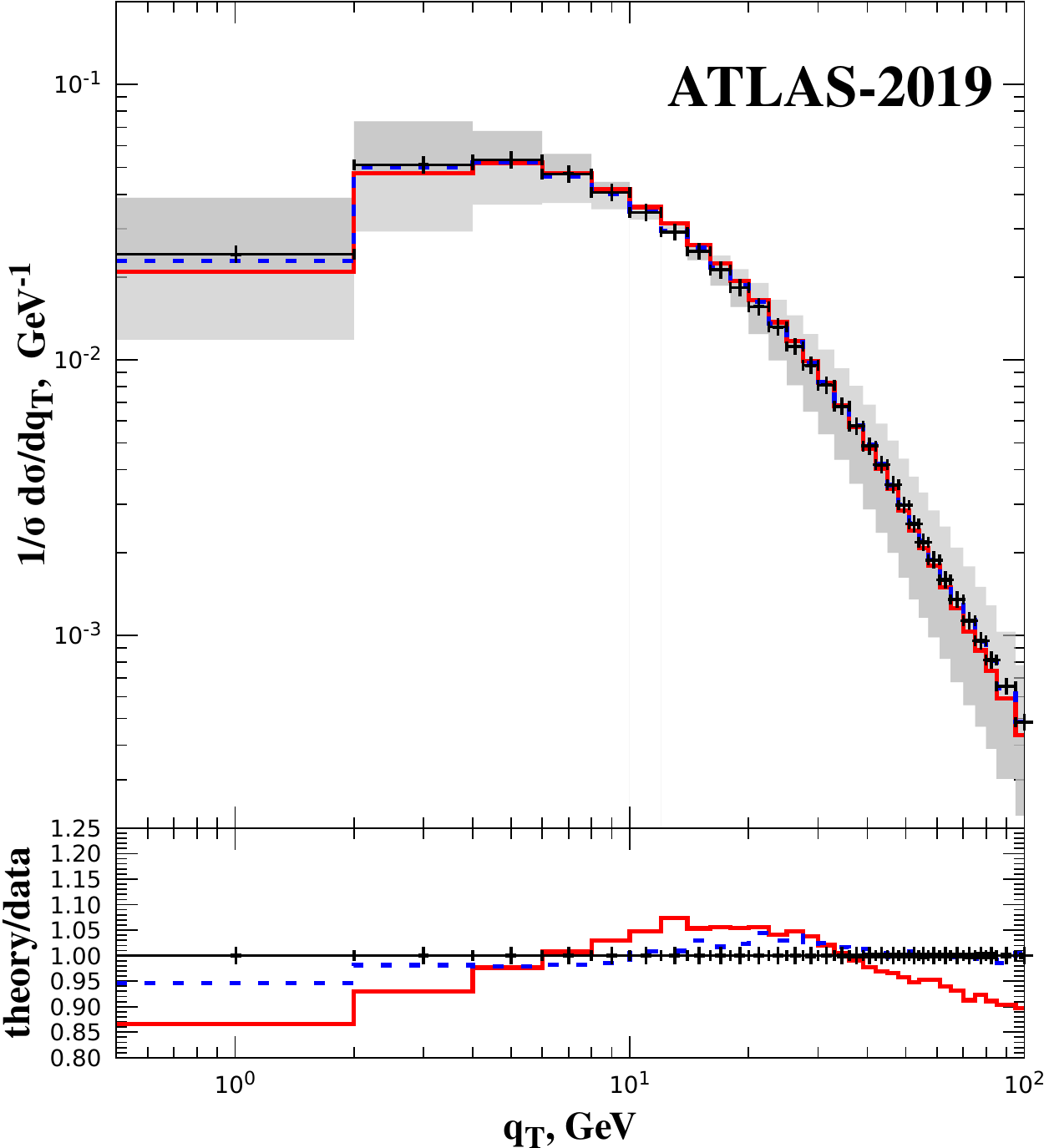}
\includegraphics[width=0.49\textwidth]{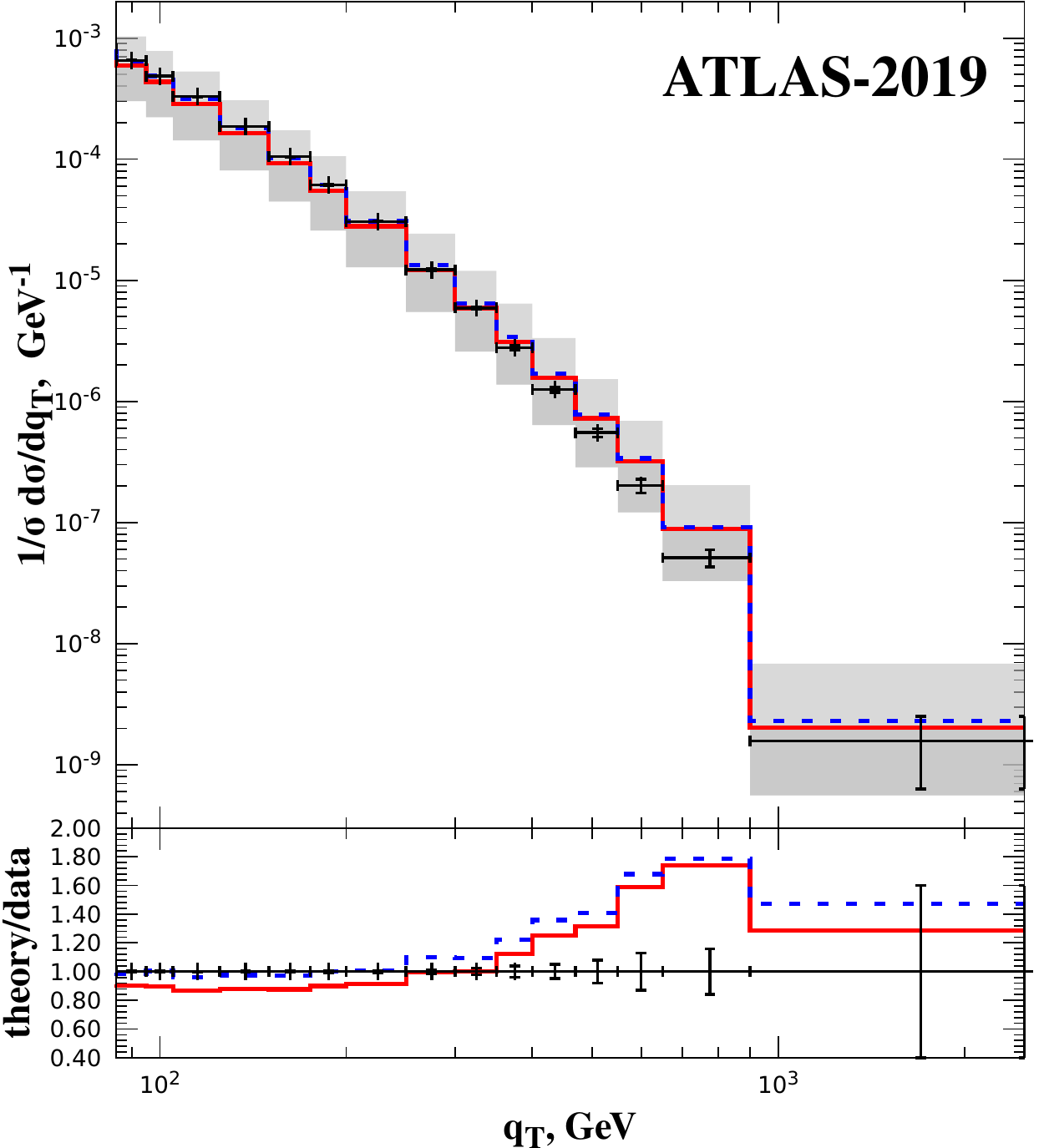}
\end{center}
\caption{The normalized transverse-momentum spectrum of Drell-Yan lepton pairs measured by the ATLAS collaboration~\cite{ATLAS-2019} compared to LO PRA predictions made with LO (blue histogram) and NLO (red histogram) UPDFs. Left panel -- $|{\bf q}_T|<100$ GeV, right panel -- $|{\bf q}_T|>100$ GeV. The uncertainty band is shown only for the LO prediction. \label{fig:Z-ATLAS}}
\end{figure}

In the Fig.~\ref{fig:Z-ATLAS} we compare our predictions for $(1/\sigma)d\sigma/d|{\bf q}_T|$-spectrum with very recent experimental data obtained by ATLAS Collaboration at CERN LHC in $pp$-collisions with $\sqrt{S}=13$ TeV~\cite{ATLAS-2019}. The same range of dilepton invariant masses as in the CDF measurement had been used by ATLAS Collaboration, while fiducial phase-space of the ATLAS measurement have covered lepton rapidities $|y_l|<2.5$ and $p_T^{(l)}>27$ GeV. For the total cross-section of dilepton production in $pp$-collisions at $\sqrt{S}=13$ TeV we have found a smaller $K-$factor than in the CDF case (Tab.~\ref{tab:data}). Our description of the shape of $|{\bf q}_T|$-distribution at higher energy (Fig.~\ref{fig:Z-ATLAS}) is significantly better than in the CDF-case(Fig.~\ref{fig:Z-CDF}), with the NLO UPDF result being clearly improved compared to the LO UPDF prediction. The central prediction with the NLO UPDF describes ATLAS data essentially within experimental uncertainties all the way up to $|{\bf q}_T|/M_Z\simeq 2.7$ which corresponds to $|{\bf q}_T|/\sqrt{S}\simeq 0.02$. The latter number is consistent with what we have obtained above in the CDF case, thus we conclude, that accuracy of our approximation is indeed controlled not by $|{\bf q}_T|/Q$ as in standard TMD-factorization, but by the ratio of characteristic scale of the process to $\sqrt{S}$ or equivalently by values of $x_{\pm}$. At higher transverse momenta the power-corrections w.r.t. $x_{\pm}$ become important, which can be taken into account only by  the complete NLO calculation in PRA.  In conjunction with this results we can also point towards the recent study~\cite{Blanco:2019qbm}, where UPDFs defined by Eq.~(\ref{eq:UPDF-deriv-form}) with $x$-independent Sudakov formfactor and gauge-invariant Matrix Elements with off-shell initial-state partons derived in a formalism, which is equivalent to ours~\cite{vanHameren:2016kkz,vanHameren:2019puc},  had been successfully used to describe the $Z$-boson production in proton-lead collisions.   

\begin{figure}
\begin{center}
\includegraphics[width=0.49\textwidth]{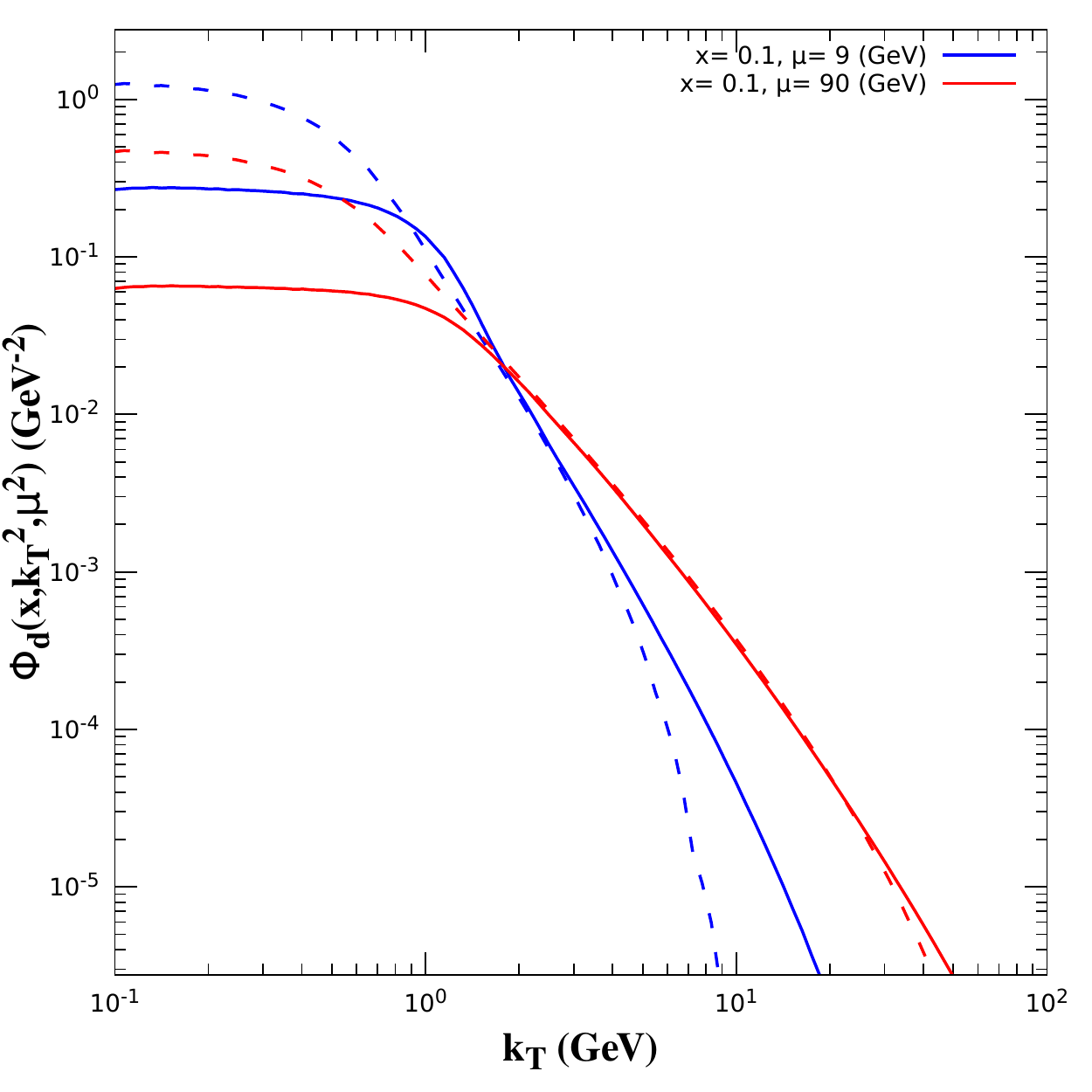}
\includegraphics[width=0.49\textwidth]{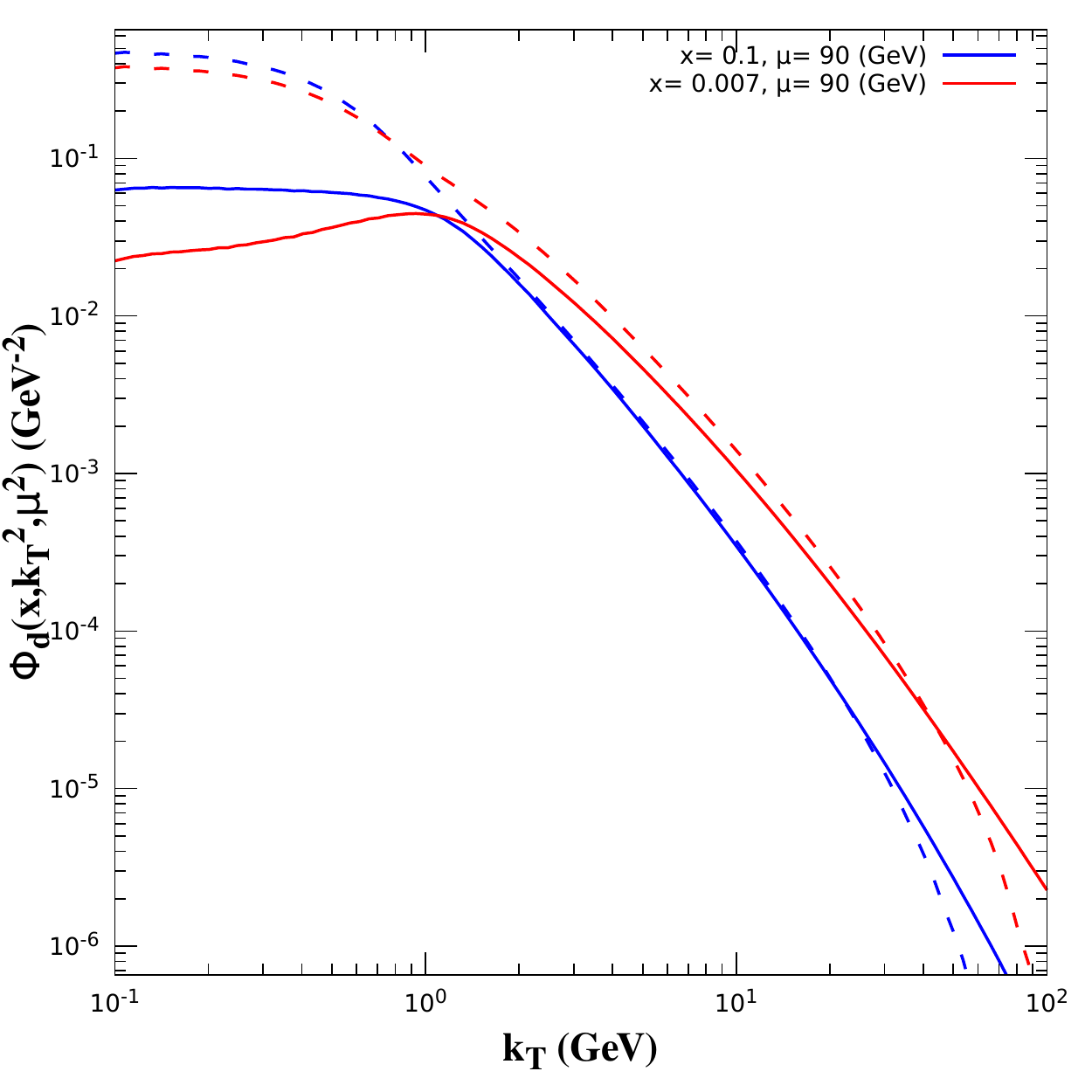}
\end{center}
\caption{Comparison of the $d$-quark NLO UPDFs (solid lines) with corresponding PB UPDFs {\tt PB-NLO-HERAI+II-2018-set1}~\cite{Martinez:2018jxt} (dashed lines). The PB UPDFs are multiplied by a factor 1.5. Left panel demonstrates $\mu$-dependence, right panel -- $x$-dependence of the $|{\bf q}_T|$-distribution. \label{fig:PB-comp}}
\end{figure}

 In the Fig.~\ref{fig:PB-comp} we compare our quark UPDFs with UPDFs obtained in the receltly-proposed Parton-Branching(PB) method~\cite{Martinez:2018jxt, Hautmann:2019biw}. The latter UPDFs can be obtained from the {\tt TMDlib} package~\cite{Hautmann:TMDlib}. The UPDFs in PB-method are derived as Monte-Carlo solution of a system of evolution equations constructed in such a way, that ${\bf q}_T$-integrated UPDF satisfies usual DGLAP equations, while transverse-momentum dependence of UPDF is essentially determined from the ambiguity in definition of ``non-resolved'' parton branchings by means of a suitable cutoff function and several scale-choices in the definition of Sudakov formfactor and branching probability. From the Fig.~\ref{fig:PB-comp} one observes, that our NLO UPDFs for $d$-quark have essentially the same shape in the region 1 GeV$<|{\bf q}_T|<\mu$ as PB UPDFs, while the latter are different from ours in overall normalization, which could be partially explained by the fact, that PB UPDFs use another  PDF set -- {\tt HERAPDF20-NLO-EIG}~\protect\cite{Abramowicz:2015mha,Buckley:2014ana} as a collinear input. The same relation between two UPDFs in the region 1 GeV$<|{\bf q}_T|<\mu$ can be found for other flavors. However, two UPDFs are dramatically different for $|{\bf q}_T|\gtrsim \mu$, where our NLO UPDF (as well as LO one) exhibits a power-like tail, while PB UPDF drops exponentially. This difference is extremely important for the description of the region of Drell-Yan spectrum with $|{\bf q}_T|\sim Q$, where LO calculation with PB UPDFs will significantly under-estimate the cross-section. To overcome this problem, authors of recent Ref.~\cite{Martinez:2020fzs} attempt to match the LO calculation with PB UPDFs with NLO QCD corrections obtained via the standard implementation of the {\tt MC@NLO} method~\cite{Alwall:2014hca}. In such a way, satisfactory description of shapes and normalization of low-energy Drell-Yan data, as well as ATLAS data on $Z$-boson ${\bf q}_T$-spectrum for $|{\bf q}_T|<10$ GeV has been obtained in Ref.~\cite{Martinez:2020fzs}. However the standard {\tt MC@NLO} method is not designed to properly take into account the off-shell initial state partons and hence it's application together with UPDFs is hard to justify. The consistent formalism of NLO calculations with off-shell initial-state partons is currently under development~\cite{vanHameren:2017hxx,Blanco:2020akb,NS_DIS1,gaQq-real-photon,Nefedov:2019mrg,Nefedov:2020ecb}. Moreover, taking into account the power-like tail of UPDF, together with effects of quark Reggeization, allows one to extend the range of applicability of High-Energy Factorization for $Z$-boson production at $\sqrt{S}=13$ TeV all the way up to $|{\bf q}_T|\lesssim 200$ GeV as we have shown above.     

 In the recent Ref.~\cite{Golec-Biernat:2019scr} the gluon and quark UPDFs have been constructed as a solutions of CCFM-Kwieci\'nski evolution equation, and consistency of this formalism with CSS-formalism up to NLL-approximation has been demonstrated. A good description of shapes of Drell-Yan $|{\bf q}_T|$-spectra at low energies, similar to ours (Figs.~\ref{fig:fit}--\ref{fig:fit2}), has been obtained in Ref.~\cite{Golec-Biernat:2019scr}. It is interesting to note, that evolution equations in this formalism are very similar to the PB evolution equations, while the logic to obtain them is different. It is tempting to suggest, that closed-form solution of CCFM-K or PB evolution equations could be found in terms of underlying collinear PDFs,  analogous to our Eqns.~(\ref{eq:UPDF}), (\ref{eq:Sudakov}), (\ref{eq:tau}) and (\ref{eq:dtau}).

 Finally, we shall discuss how our approach describes the angular distribution of leptons in the rest-frame of the lepton-pair, which can be parametrised as follows:
\begin{eqnarray}
\frac{d\sigma}{dQ d{\bf q}_T^2 dy d\Omega_l} &=& \frac{3}{16\pi} \frac{d\sigma}{dQ d{\bf q}_T^2 dy} \left\{ (1+\cos^2\theta_l) +\frac{A_0}{2}(1-3\cos^2\theta_l) \right. \nonumber \\
&+& A_1\sin2\theta_l \cos\phi_l + \frac{A_2}{2}\sin^2\theta_l\sin 2\phi_l + A_3\sin\theta_l\cos\phi_l +A_4\cos\theta_l \nonumber \\
&+& \left. A_5 \sin^2\theta_l \sin 2\phi_l + A_6 \sin 2\theta_l \sin\phi_l + A_7\sin\theta_l\sin\phi_l \right\} ,
\end{eqnarray}
where angular coefficients $A_i$ are functions of $Q^2$, ${\bf q}_T^2$ and $y$, realted with the polarization density-matrix of the intermidiate vector-boson. Thus the study of transverse-momentum dependence of angular coefficients will allow us to check whether the spin structure of our MMRK amplitudes can reasonably approximate the spin structure of exact QCD amplitudes. 

To obtain theoretical predictions for angular coefficients with the help of our master-formula for the cross-section~(\ref{eq:CS_2-2_PRA-MASTER}), we use the same harmonic-projectors method which has been used to obtain theoretical predictions in the Ref.~\cite{ATLAS-2016}, see Eqns. (4) and (5) of the the Sec. 2 in this reference. 

\begin{figure}
\begin{center}
\includegraphics[width=0.49\textwidth]{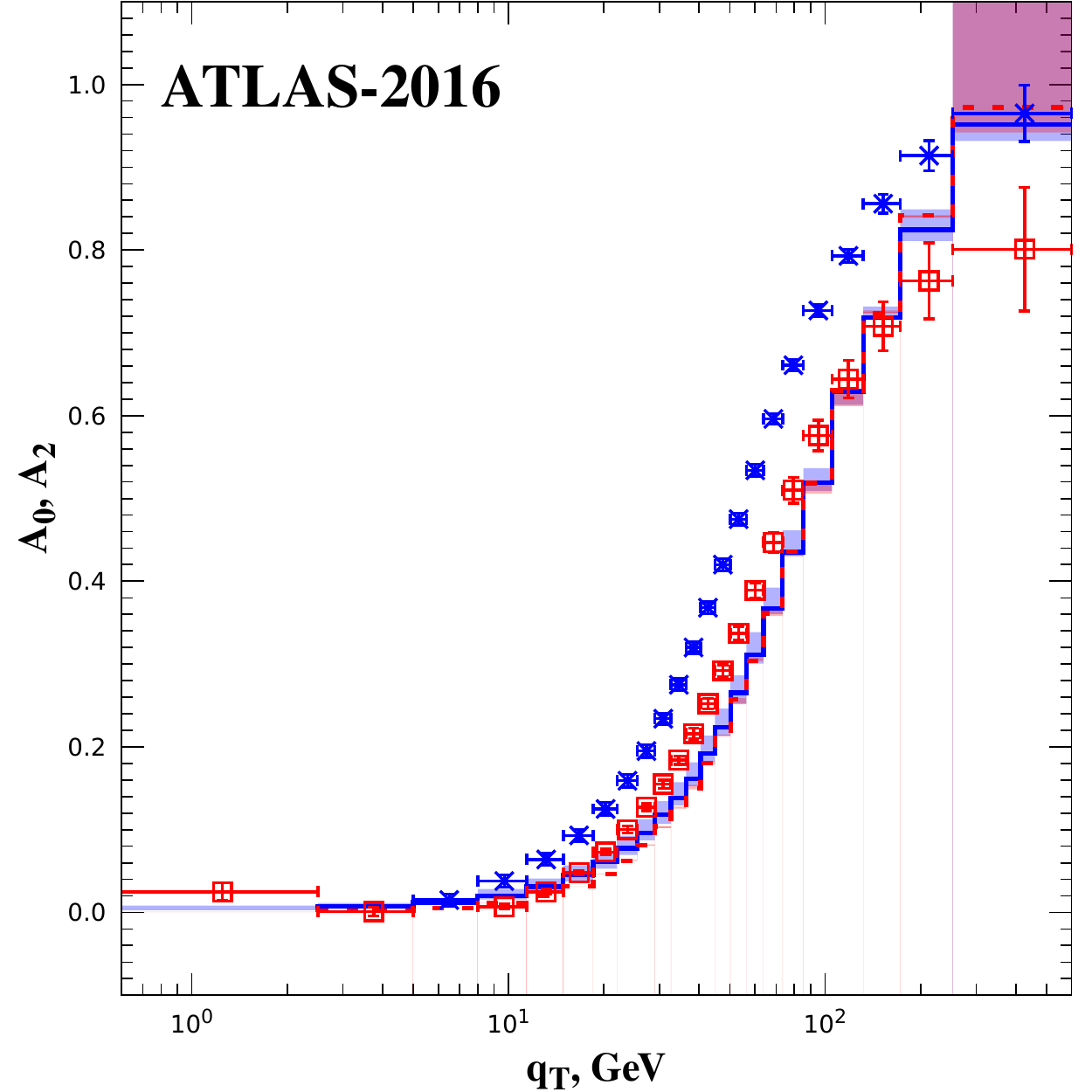}
\includegraphics[width=0.49\textwidth]{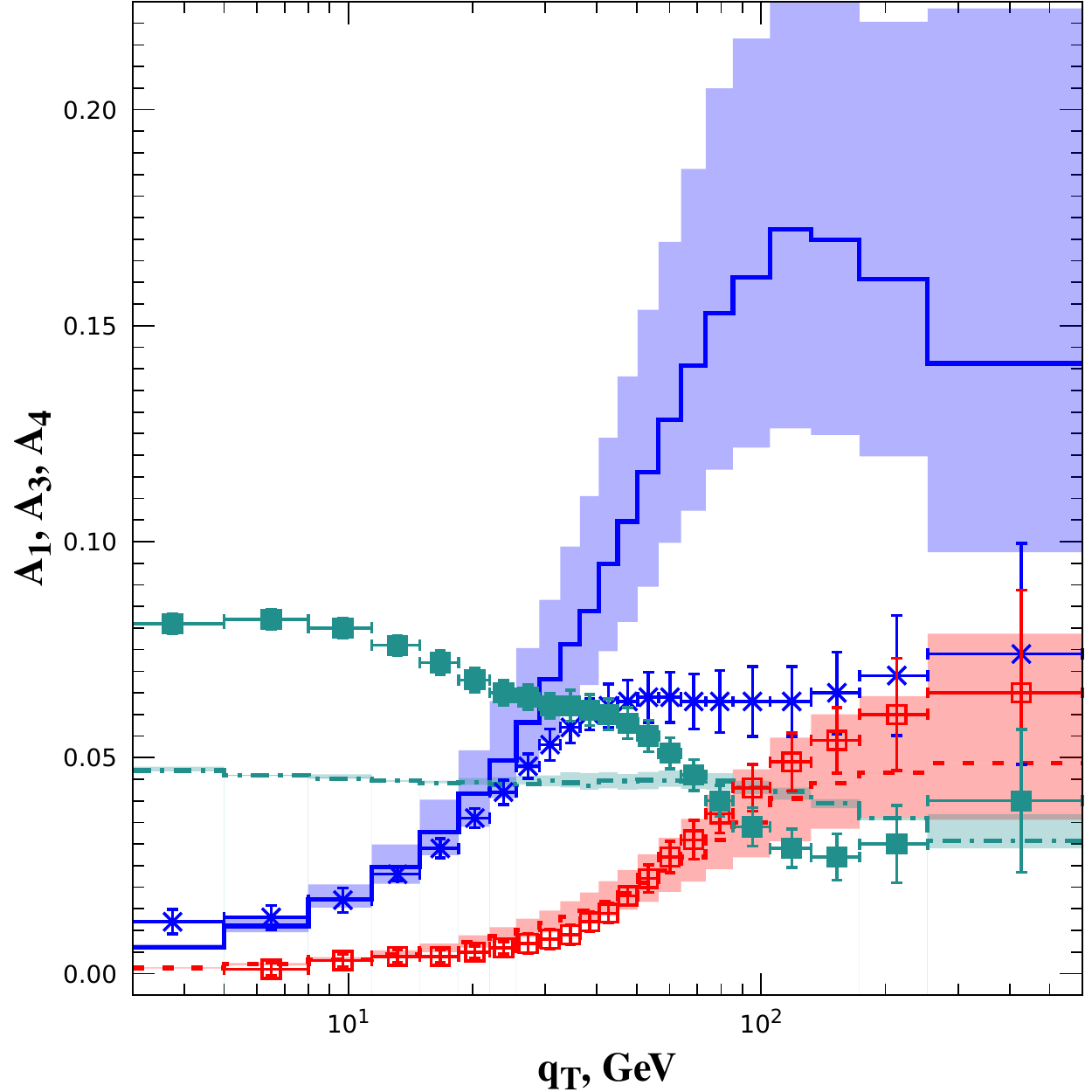}
\end{center}
\caption{Comparison of angular coefficients which are nonzero in the LO of PRA with corresponding experimental data obtained by ATLAS Collaboration~\cite{ATLAS-2016}. Left panel: solid histogram and data-points marked by crosses -- $A_0$, dashed histogram and boxes -- $A_2$. Right panel: solid histogram and data-points marked by crosses -- $A_1$, dashed histogram and open boxes -- $A_3$, dash-dotted histogram and filled boxes -- $A_4$. \label{fig:ATLAS_A0-A4}}
\end{figure}

  In the Fig.~\ref{fig:ATLAS_A0-A4} we compare our theoretical predictions for angular coefficients with experimental data obtained by ATLAS Collaboration~\cite{ATLAS-2016} in $pp$-collisions with $\sqrt{S}=8$ TeV in the range of lepton-pair invariant masses $80<Q<100$ GeV and lepton-pair rapidities $|y|<2$. The results on coefficients $A_0$ and $A_2$ are especially interesting since in the CPM the Lam-Tung relation~\cite{Lam:1978zr, Lam:1978pu}: $A_0=A_2$ holds up to NLO in $\alpha_s$ and is violated only by NNLO effects. Experimentally, $O(10\%)$-violation of Lam-Tung relation is observed at high-$|{\bf q}_T|$ and high energy, however LO of PRA predicts much smaller $O(1\%)$-violation at $Q\simeq M_Z$, see the left panel of the Fig.~\ref{fig:ATLAS_A0-A4}. More significant violation of Lam-Tung relation is predicted~\cite{Nefedov:2012cq} by LO of PRA for smaller values of $Q$, and this effect increases with increasing $\sqrt{S}$ -- a trend worth to be studied experimentally. Very similar results for coefficients $A_0$ and $A_2$ where found in the LO PRA calculations of Ref.~\cite{Motyka:2016lta} with different quark UPDF, see Fig. 9(d) in this reference.

  The coefficients $A_1$, $A_3$ and $A_4$ are nonzero only if parity-violating couplings of $Z$-boson are present, see the right panel of Fig.~\ref{fig:ATLAS_A0-A4}, while coefficients $A_5$, $A_6$ and $A_7$ further require effects beyond NLO of CPM to be taken into account and these coefficients are zero in the LO of PRA. From the right panel of Fig.~\ref{fig:ATLAS_A0-A4} one observes, that ${\bf q}_T$-dependence of $A_3$ is nicely described in the LO of PRA, while $A_1$ and $A_4$ come-out to be of the right order of magnitude and roughly of a correct shape. Nevertheless, it is clear, that LO of PRA is not capable to correctly capture this subtle details of a spin structure of production amplitude, and full NLO corrections in PRA, which will exactly take into account emission of an additional parton from the hard process, are necessary to quantitatively predict all angular coefficients.   

\section{Conclusions and outlook}
\label{sec:concl}
In the present paper we have introduced a new prescription to obtain UPDF from LO and NLO collinear PDFs and to define it's non-perturbative ambiguity. This UPDF, together with QCD and QED gauge-invariant matrix elements already in the LO in $\alpha_s$ provide an excellent description of shapes of transverse-momentum distributions of DY lepton-pairs in  $pp$ and $p\bar{p}$-collisions at low and high collision energies, in the region $Q_T/\sqrt{S}\ll 1$. Qualitative description of transverse-momentum dependence of angular coefficients of lepton distribution in the rest frame of the lepton pair is also achieved for $Q\simeq M_Z$. To describe the normalization of $|{\bf q}_T|$-distribution and extend the formalism to higher values of transverse momenta it is necessary to go beyond LO in $\alpha_s$ for the coefficient function in our approach. The formalism of NLO calculations is currently under development~\cite{vanHameren:2017hxx,Blanco:2020akb,NS_DIS1,gaQq-real-photon,Nefedov:2019mrg,Nefedov:2020ecb}.     

\section*{Acknowledgments}

 Authors are grateful to Prof. B.A.Kniehl and Dr. Zhi-Guo He for enlightening discussions on the CSS formalism, which served as motivation for the present work and for computational resources provided by the II Institute for Theoretical Physics of Hamburg University.  The work has been supported in parts by the Ministry of Education and Science of Russia via State assignment to educational and research institutions under project FSSS-2020-0014 and by the Foundation for the Advancement of Theoretical Physics and Mathematics BASIS, grant No. 18-1-1-30-1.

\bibliography{mybibfile,DY-data}

\end{document}